\documentclass[aps,prl,reprint,superscriptaddress,nofootinbib,floatfix]{revtex4-2}
\usepackage[T1]{fontenc}
\usepackage{amsmath,amssymb,bm,graphicx,mathptmx}
\usepackage[hidelinks]{hyperref}
\newcommand{\Tr}{\operatorname{Tr}}
\newcommand{\e}{\mathrm e}
\begin{document}
\title{Detector Placement Relaxes Efficiency Requirements for Detecting a Weak Coherent Link}
\author{M. H. Yu}
\affiliation{College of Physics and Materials Science, Tianjin Normal University, Tianjin 300387, China}
\author{S. Y. Li}
\affiliation{College of Physics and Materials Science, Tianjin Normal University, Tianjin 300387, China}
\author{X. Z. Zhang}
\email{zhangxz@tjnu.edu.cn}
\affiliation{College of Physics and Materials Science, Tianjin Normal University, Tianjin 300387, China}
\affiliation{Interdisciplinary Center, Tianjin Normal University, Tianjin 300387, China}
\begin{abstract}
A weak coherent link can produce unusually early arrivals without changing stationary count rates. Missed jumps along routes available without the link can produce similar records, obscuring its presence. On an eight-site ring, relocating the same detector losses exposes one more crossing on the easiest background route to hide, without changing the dynamics or leading link signal. At fixed error and a common observation-time scaling, the tolerated missed fraction and time-bin width follow fourth-root rather than square-root scaling with link strength. These requirements hold for every decision using the counting record and are met using three event counts per bin, with all elapsed time included. At the same stationary collected flux, where jumps are monitored can therefore determine the time resolution needed for weak-link sensing.
\end{abstract}
\maketitle

\textit{Introduction.---}Quantum sensing uses detected events to infer weak
interactions~\cite{Degen2017,Giovannetti2011,Braunstein1994}. Continuous monitoring provides
their times and types, from which count rates and temporal correlations can be
determined~\cite{Dalibard1992,Plenio1998,Mabuchi1996,Gammelmark2013,Kiilerich2014,Gammelmark2014}.
A weak coherent link opens a shortcut between distant sites. An early arrival, however, does not
by itself establish that the link is present. A particle following a route available without the link can leave a similar record if its intervening jumps go undetected. The link can change arrival times while leaving stationary count rates unchanged.
Can detector placement reduce the required
	efficiency and time resolution to distinguish these processes, without increasing the stationary
collected flux?

A weak signal can become easier to distinguish when it produces outcomes that the background almost never
	produces~\cite{Tsang2016,Nair2016,Paur2018,Paur2019,Darekar2026}. Detector crosstalk can remove
this advantage~\cite{Gessner2020}, while recording events at finite resolution can reduce the
information in correlated records~\cite{Villard2011}. Quantum-trajectory and waiting-time methods
already account for missed jumps, multiple detection channels, and event
correlations~\cite{Kiilerich2015,Landi2024,Radaelli2026,OConnor2024,Radaelli2023}. The remaining
question is how the location of losses sets the efficiency and time resolution needed to detect a
weak transport link. An excess of early arrivals alone cannot settle this question. Other events and correlations may carry information, and collecting further events takes time. Any bound on the observation time must therefore include the
complete record and all elapsed time.

We compare two detector arrangements, C and S, on a fixed eight-site ring. In S,
the easiest background route to hide crosses one more well-monitored bond than in C, while the leading direct-link signal is
unchanged. The weaker background leaves longer waits informative and permits coarser timing. At a common observation-time scaling, the tolerated missed fraction and bin width change from square-root order in C to fourth-root order in S as the link weakens. We bound the time required
by every decision using the binned counting record. A rule using three event counts per bin attains
the same powers. At common settings, the dynamics, stationary collected flux, and entire selected return-time distribution are identical. The
advantage thus comes from revealing background routes, rather than producing sensing opportunities more rapidly.

\begin{figure*}[t]
	\centering\includegraphics[width=\textwidth]{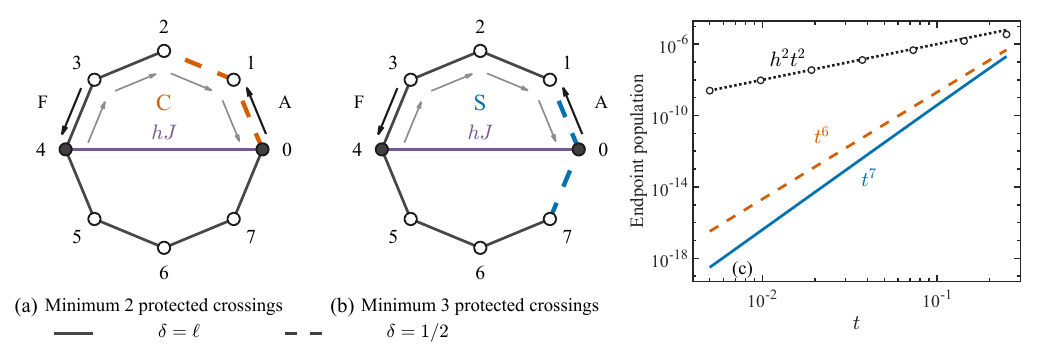}
\caption{Relocating losses suppresses background arrivals at site 0 with no intervening detection while preserving the leading link signal.
	(a) and (b) Ring bonds have coherent hopping $J$ and jump rate $\gamma J$ in each
	direction; the 4--0 link has hopping $hJ$. Solid bonds are well monitored (protected), with missed
	fraction $\ell$; dashed bonds have missed fraction $1/2$. The two routes from 4 to 0 cross two and four
	protected bonds respectively in C, but three each in S. Gray arrows trace one background route;
	black arrows mark the selected event $F=(3\to4)$ and target $A=(0\to1)$.
	(c) Unnormalized populations at site 0 after preparation at site 4, with no intervening detection.
	Orange dashed and blue solid curves show the null populations $F_{0,0}^{\rm C}(0,4;t)$ and
	$F_{0,0}^{\rm S}(0,4;t)$. The curves are calculated by matrix exponentiation at $\gamma=1$, $\ell=0$ on logarithmic axes; the analytical short-time powers are $t^6$ and $t^7$. Black circles show the S signal increment $F_{h,0}^{\rm S}(0,4;t)-F_{0,0}^{\rm S}(0,4;t)$ at
	$h=0.01$; the dotted line is $h^2t^2$. Time is $t=Jt_{\rm phys}$. These local probabilities explain the
	mechanism; the text and Supplemental Material (SM) establish the observation-time bounds.}
	\label{fig:routes}
\end{figure*}

\textit{Model and detected events.---}A particle moves on eight sites $j=0,\ldots,7$, with indices
taken modulo eight. The state localized at site $j$ is $P_j=|j\rangle\langle j|$. Frequencies
are in units of the fixed ring hopping $J>0$, and times in $J^{-1}$. The Hamiltonian and jump
operators are
\begin{align}
	H_h&=\sum_{j=0}^7(|j+1\rangle\langle j|+\mathrm{H.c.})
	+h(|0\rangle\langle4|+\mathrm{H.c.}),\label{eq:H}\\
	L_{j,+}&=\sqrt\gamma|j+1\rangle\langle j|,\qquad
	L_{j,-}=\sqrt\gamma|j\rangle\langle j+1|.\label{eq:L}
\end{align}
Here $h$ is the dimensionless link strength. The coherent hoppings are real, and the jump
rate $\gamma>0$ is fixed and known. Jumps occur on ring bonds; the 4--0 link is purely coherent.
At each $h$, both detector arrangements have the same Lindblad generator
$\mathcal L_h$~\cite{Lindblad1976,Gorini1976}. A jump in channel $\mu$ is missed with probability
$\delta_\mu$. Arrangement C misses half the jumps in both directions on bonds $0$--$1$ and
$1$--$2$; S does so on $7$--$0$ and $0$--$1$. The other twelve directed channels share the missed
fraction $\ell$. We call their six bonds protected: a jump crossing one is detected with
probability $1-\ell$. Protection refers to observation, not a barrier to motion.

The particle is prepared once in $P_4$ to distinguish $h=0$ from a specified $h>0$. Efficiencies
are known; detector settings and the test may be chosen for this target $h$, then held fixed under
both hypotheses. Starting at preparation, a clock divides absolute time into bins of width
$\Delta$. Each bin records the numbers of events in three classes,
\begin{equation}
	F=(3\to4),\quad A=(0\to1),\quad O=\text{other registered events}.
	\label{eq:classes}
\end{equation}
An $F$ detection places the particle at site 4. An $A$ detection reports occupation of site 0
immediately before the jump. Every $O$ event is counted, but its direction is discarded. Within
each bin, neither event times nor their order are retained. Observation ends at a predetermined
time; empty bins and intervals that yield no useful detection still count towards its duration. There are no external resets, feedback, dark counts, dead time, or label
errors. A false alarm declares a link under $h=0$; a missed link declares none under $h>0$. For
arrangement $D$, let $T_e^D(h,\ell,\Delta)$ be the shortest time at which both probabilities can be
at most a fixed $e\in(0,1/2)$. Using their equal-weight average as the error criterion gives the same time
	scalings. The optimization covers this counting experiment, not arbitrary output measurements.

When no event is detected, the particle can still undergo missed jumps. Its conditional state, weighted by the probability of no detection, therefore evolves under
\begin{equation}
	\mathcal K_{h,\ell}(X)=-i[H_h,X]-2\gamma X
	+\sum_\mu\delta_\mu L_\mu X L_\mu^\dagger.\label{eq:K}
\end{equation}
For a particle starting at $b$, define $F_{h,\ell}(a,b;t)=\langle a|\e^{t\mathcal K_{h,\ell}}P_b|a\rangle$.
This is the unnormalized population at $a$ with no intervening detection. Its sum over $a$ is the
probability of no detection before $t$. The probability density that the next detected jump is
$a\to c$ is $\gamma(1-\delta_{ac})F_{h,\ell}(a,b;t)$. This jump leaves the particle in
$P_c$~\cite{Carmichael1993,Wiseman2010,Bouten2007}.

The physical stationary state is $\pi=I/8$ for every real $h$. Each directed channel therefore
carries flux $\gamma/8$, and both detectors register the fraction
\begin{equation}
	q_{\rm coll}=\frac78-\frac{3\ell}{4}.\label{eq:collection}
\end{equation}
The stationary registered-event rate is $2\gamma q_{\rm coll}$. Total collection approaches $7/8$ as
$\ell\to0$; only the protected channels approach perfect efficiency. Each channel's stationary rate
is independent of $h$. At $h=0$, a rotation of the ring by one site also relates the generators $\mathcal K_{0,\ell}$ of C and S, giving identical spectra.

To find the times between $F$ detections, sum over all intermediate outcomes. The resulting
generator is $\mathcal Q_h=\mathcal L_h-\mathcal J_F$, where $\mathcal J_F(X)=\gamma(1-\ell)X_{33}P_4$.
It gives $-\mathcal Q_h^{-1}P_4=I/[\gamma(1-\ell)]$ and mean return time
$8/[\gamma(1-\ell)]$. Because $\mathcal Q_h$ is common to C and S, their entire $F$ return-time distributions
match. The detector still records the intermediate events summed over in this return law. The mean counts during the transient that follows the initial preparation can differ
and are included in the time bounds (SM~\cite{SM}, Secs.~S1 and S4). Here placement changes which
jumps are reported; control of collective order~\cite{Zhang2020PRB,Zhang2026PRB} and
relaxation~\cite{Li2026} instead changes the dynamics.

\textit{Background paths and the weak link.---}The background is the population at site 0 under $h=0$,
starting at site 4 with no intervening detection. A detected jump excludes a history from this
population; it does not stop the particle. Expanding Eq.~\eqref{eq:K} order by order in the number of missed jumps gives positive
contributions from histories containing coherent motion between those jumps. Within each coherent
segment, amplitudes are summed before taking their squared modulus. The distance dependence of coherent and stochastic short-time powers, open-system path shortening, and explicit remainder bounds are established~\cite{Szigeti2019}. A coherent
segment of $d$ bonds contributes population of order $t^{2d}$; each missed jump adds a time
integral. Coherent passage across a protected bond contributes a short-time weight of order $t^2$.
An undetected jump across the same bond contributes a weight of order $\gamma\ell t$. From site 4
to site 0,
\begin{equation}
	F_{0,\ell}^{\rm C}(0,4;t)\asymp t^4(\ell+t)^2,
	\qquad F_{0,\ell}^{\rm S}(0,4;t)\asymp t^4(\ell+t)^3.
	\label{eq:backgrounds}
\end{equation}

Here $\asymp$ means equality up to positive constant factors, uniformly at short times and small
$\ell$. In C, the easiest four-bond background route to hide crosses two half-loss bonds and two protected bonds. Either route in S crosses three
protected bonds. At $\ell=0$, protected bonds must be crossed coherently to avoid detection. The
leading C route combines two missed jumps and two coherent crossings, giving $t^2t^4=t^6$. Each
leading S route combines one missed jump and three coherent crossings, giving $tt^6=t^7$. The ring
distance remains four [Fig.~\ref{fig:routes}]. For nonzero $\ell$, the S background gains an additional factor $\ell+t$. Coherent motion across protected bonds
dominates for $t\gg\ell$, and missed jumps for $t\ll\ell$, at fixed $\gamma$. The calculation
retains interference between coherent paths; bond counting alone does not establish these powers.

The weak link bypasses these routes. The leading contribution of
	the link is the same in both arrangements,
\begin{equation}
	F_{h,\ell}(0,4;t)-F_{0,\ell}(0,4;t)=h^2t^2[1+O(t)].
	\label{eq:signal}
\end{equation}
Reversing the sign of the real link leaves all counting probabilities unchanged, so odd powers of $h$ vanish (SM, Sec.~S2.1). Equation~\eqref{eq:signal} gives the nonzero quadratic term. At $\ell=0$, balancing $t^6$ or $t^7$ against $h^2t^2$ gives waits
$t_{\rm C}\asymp h^{1/2}$ and $t_{\rm S}\asymp h^{2/5}$, at fixed $\gamma$. Thus S leaves longer waits
informative as $h$ decreases. This explains the mechanism; establishing the required clock resolution
also requires the complete record.

A detected $F$ jump places the particle at site 4. We compare the waiting-time densities for $A$ to
be the next detection, with and without the link. Integrating the squared difference between their
square roots gives the target-channel contribution to the squared Hellinger distance. This distance
measures statistical distinguishability. Write $p$ for the minimum number of protected bonds on a background route from site 4 to site 0; $p=2$ for C and $p=3$ for S. With a short-time cutoff $\tau>0$ independent of $h,\ell$,
rationalizing the square-root difference gives the scaling
\begin{equation}
	h^4\int_0^\tau\frac{t^2\,dt}{t^2(\ell+t)^p+h^2}
	\asymp \frac{h^4}{(\ell+h^{2/(p+2)})^{p-1}}.\label{eq:localrate}
\end{equation}
Both background and signal enter the denominator: at the shortest waits, the signal can exceed the
background although both probabilities are small. Waits of order $\ell+h^{2/(p+2)}$ dominate the
	integral. Their information comes from intervals with no detection, without an enhanced spectral response~\cite{Lau2018,Langbein2018}.
With exact timestamps, the finite mean return time supplies sensing opportunities at a rate bounded away
from zero. Equation~\eqref{eq:localrate} already accounts for how rarely a useful wait occurs in each such opportunity.

For $\ell>h^{2/(p+2)}$, waits of order $\ell$ give information of orders $h^4/\ell$ for C and
$h^4/\ell^2$ for S. A fixed nonzero $\ell$ therefore restores $h^{-4}$ observation-time scaling. S
tolerates more loss because its background must cross one additional protected bond without detection;
the direct link bypasses that bond.

\begin{figure*}[t]
	\centering\includegraphics[width=\textwidth]{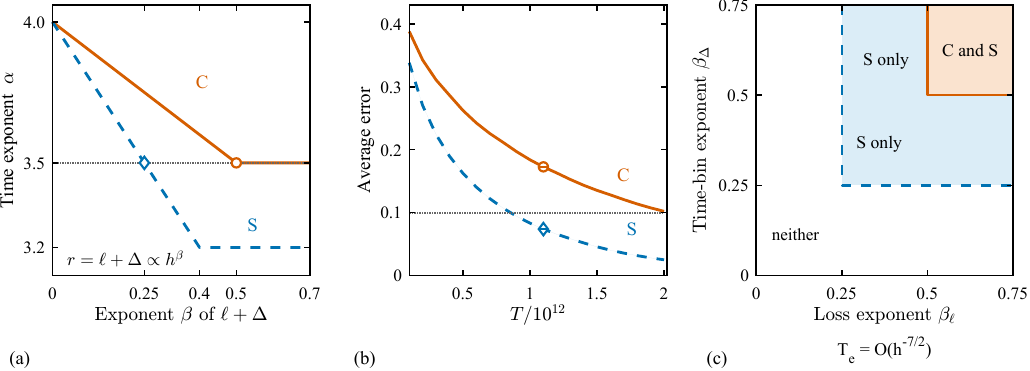}
\caption{Detector placement changes the efficiency and timing required at a common observation-time scaling.
	(a) Exponents in $T_e=\Theta(h^{-\alpha})$ for $\ell+\Delta=\kappa h^\beta$, with fixed $\kappa,\gamma>0$:
	$\alpha_{\rm C}=4-\min(\beta,1/2)$ and $\alpha_{\rm S}=4-2\min(\beta,2/5)$. Parameters remain in the
	small-loss, small-bin domain of the theorem, including at $\beta=0$. The dotted line marks the reference scale $h^{-7/2}$.
	(b) Bounds on the minimum average error $R_D^{\rm count}$, with equal hypothesis weights, from the total
	count of $(F,0,0,0,A)$. Each arrangement has a separately calibrated decision. Parameters are $\gamma=1$,
	$h=10^{-3}$, $\ell=10^{-4}$, and $\Delta=0.04$. Bands enclose the minimum error for the correlated count at
	twenty observation times; lines guide the eye. Band widths are smaller than line thickness. Markers indicate the time in Eq.~\eqref{eq:finite}, including the initial wait of 20 and every block. These bounds concern this count, not arbitrary processing of the
	full record.
	(c) Allowed exponents at time scale $h^{-7/2}$ for $\ell=\kappa_\ell h^{\beta_\ell}$ and
	$\Delta=\kappa_\Delta h^{\beta_\Delta}$. S requires both exponents at least $1/4$; C requires at least
	$1/2$, including boundaries. Smaller exponents allow more missed jumps and wider bins. Panels (a,c) follow
	from Eqs.~\eqref{eq:timeC}--\eqref{eq:timeS}; (b) uses matrix exponentials with explicit error bounds and
	a finite-time Poisson comparison (SM, Secs.~S7 and S8). Times and bin widths are in $J^{-1}$.}
	\label{fig:resources}
\end{figure*}

\textit{Efficiency and time resolution.---}The minimum observation times depend on both missed jumps and the
detector clock. At fixed $\gamma$ and error bound $e$, for sufficiently small $h>0$,
$0\le\ell\le\ell_0<1/2$, and $0<\Delta\le\Delta_0$, they obey
\begin{align}
	T_e^{\rm C}(h,\ell,\Delta)&=\Theta\!\left[h^{-4}(\ell+\Delta+\sqrt h)\right],\label{eq:timeC}\\
	T_e^{\rm S}(h,\ell,\Delta)&=\Theta\!\left[h^{-4}(\ell+\Delta+h^{2/5})^2\right].\label{eq:timeS}
\end{align}
The symbol $\Theta$ denotes upper and lower bounds differing by positive constant factors. These
constants are independent of $h,\ell,\Delta$ in the stated domain, but may depend on
$\gamma,e,\ell_0,\Delta_0$. They are not uniform as $\gamma$ tends to zero or infinity.

The clock bins share one absolute time axis. An event's unknown position inside its bin affects both
neighboring waits, one before and one after the event, so later
detections may help locate it.  The lower bound allows these
correlations and even retains every directed label and within-bin event order. The three counts in
Eq.~\eqref{eq:classes} attain the same time powers.

Missed jumps alter no-detection evolution; bins instead conceal the times of registered events. Two
separate bounds are therefore needed. One allows exact timestamps but includes missed jumps; the other
sets $\ell=0$ but retains finite bins. Both cover all reset sites, long waits, and the final interval without detection. To
transfer the latter bound to $\ell>0$, delete each protected-channel report independently with
probability $\ell$. The physical jump still occurs. This rule is identical under both hypotheses and
cannot improve discrimination. Deletion must precede combining labels into $O$, which mixes channels
of different efficiencies. Applying a positive-$\ell$ decision rule to this thinned zero-$\ell$ record reproduces its error probabilities, so it cannot beat the bound obtained at $\ell=0$. The larger of the two required times
gives the lower bounds in Eqs.~\eqref{eq:timeC} and \eqref{eq:timeS} (SM, Secs.~S3--S6).

A simple counting rule attains these powers. Combine an integer number of detector bins into intervals
of width $W=\Delta\max\{1,\lceil(\ell+h^{2/(p+2)})/\Delta\rceil\}$. Form nonoverlapping pairs
of these intervals. A pair is successful only if its first interval contains exactly one $F$ event and
its second contains exactly one $A$ event. Neither interval may contain any other registered event.
Every other pair is a failure. The stationary success probabilities without and with the link, $m_0$
and $m_h$, obey
\begin{equation}
	m_0\asymp W^{p+6},\qquad
	m_h-m_0=\frac{7\gamma^2(1-\ell)}{96}h^2W^4[1+O(W)].
	\label{eq:countshift}
\end{equation}
The calculation includes no-detection evolution before, between, and after the two events. The $F$ jump
leaves the particle at site 4, but subsequent missed jumps generally leave a mixed state at the pair
boundary.

Successive pairs are correlated because the particle is not reset at their boundaries. The dissipator
limits how long these correlations persist. State differences decay at rate at least
$\lambda=2\gamma[1-\cos(\pi/4)]$ in the Hilbert--Schmidt norm, uniformly in $h$. For any incoming state, the probability that
	a pair is successful is $O(W^{p+6})$. The covariance contains two such small factors, so correlations
do not change the required time power. Including the initial transient from $P_4$, a threshold midway between the two stationary count means succeeds for $n\asymp h^{-4}W^{p-2}$ pairs. All pairs, successful or not, contribute to
the time $2nW$ (End Matter and SM, Sec.~S6). The choice $W\asymp\ell+\Delta+h^{2/(p+2)}$ balances more
arrivals against more background. Calibration must resolve the difference between pattern
probabilities and may require greater precision as $h$ decreases.

At $\ell=0$ with sufficiently fine timing, C's fastest scaling is $\Theta(h^{-7/2})$. To reach the common time scale $T_e=O(h^{-7/2})$, the necessary and
sufficient conditions on $r=\ell+\Delta$ are
\begin{equation}
	r=O(h^{1/2})\quad(\mathrm C),\qquad
	r=O(h^{1/4})\quad(\mathrm S).\label{eq:resources}
\end{equation}
Each nonnegative term in $r$ must satisfy its condition. This compares allowed resources at a common
time scale. To compare observation times at equal collection and clock resolution, fix common
$\ell,\Delta$ with $\ell+\Delta=\kappa h^{1/4}$, $\kappa>0$ fixed. C then requires $\Theta(h^{-15/4})$ and
S $\Theta(h^{-7/2})$, giving $T_e^{\rm S}/T_e^{\rm C}=\Theta(h^{1/4})$. S's fastest scaling, $h^{-16/5}$,
requires $r=O(h^{2/5})$. If either $\ell$ or $\Delta$ is held at a positive value within the theorem's
domain as $h\to0$, both return to $h^{-4}$. Their prefactors may still differ; equal powers do not determine which arrangement performs better at
	finite $h$ [Fig.~\ref{fig:resources}(a,c)].

\textit{Finite observation times.---}A common counting rule also demonstrates an
advantage at finite parameters. Set $\gamma=1$, $h=10^{-3}$, $\ell=10^{-4}$, and $\Delta=0.04$, giving
the collection fraction $q_{\rm coll}=0.874925$. Prepare $P_4$ once, allow an initial wait of 20 time units, then count $(F,0,0,0,A)$ in disjoint
five-bin blocks. An $F$ or $A$ bin contains exactly one event of that class and no others; $0$ means no
registered event. Every block takes $0.2$, whether accepted or rejected. Let $Z_h$ describe an empty
bin, and $F_h,A_h$ the two single-event bins. The pattern map and stationary probability are
\begin{equation}
	B_h=A_h Z_h^3 F_h,\qquad \mu_h=\Tr B_h(I/8).
	\label{eq:fivebin}
\end{equation}
At these settings, S reduces $\mu_0$ by a factor of about $3.26$ while retaining $99.2\%$ of
$\mu_h-\mu_0$. This finite five-bin rule differs from the preceding two-interval rule.

Let $R_D^{\rm count}(T)$ be the minimum equal-weight average of false-alarm
and missed-link probabilities using only the total pattern count. Both arrangements use the same
statistic and resources, with separately calibrated thresholds. For $M=5.5\times10^{12}$ blocks and
common time $T=20+M/5$,
\begin{align}
	0.17277&<R_{\rm C}^{\rm count}(T)<0.17294,\nonumber\\
	0.07386&<R_{\rm S}^{\rm count}(T)<0.07402.
	\label{eq:finite}
\end{align}
These bounds include correlations and preparation in $P_4$ [Fig.~\ref{fig:resources}(b)]. In S, eleven
or more occurrences declare a link, with false-alarm and missed-link probabilities below $0.05325$ and
$0.09479$. In C, the lower bound excludes both errors being at most $0.1$ for any decision using this
count, since their average would then be at most $0.1$. This finite comparison is restricted to
the pattern total; Eqs.~\eqref{eq:timeC}--\eqref{eq:timeS} optimize the time powers for the entire
record.

The dimensionless duration is $T\simeq1.1\times10^{12}$. Assessing its practicality requires a physical
frequency scale and an implementation. The End Matter and SM, Secs.~S7 and S8, give the finite
probability and error bounds.

\textit{Conclusion.---}The placement of detection losses determines the time resolution needed to detect a weak
coherent link. Monitoring competing routes makes background detours harder to hide, leaving longer waits
informative without changing the dynamics or leading signal. On this ring, complete-record bounds and counting rules establish the resource difference with all elapsed time included. Which competing routes are monitored therefore helps determine the required timing precision, even at unchanged stationary collection.

\begin{samepage}
\begin{acknowledgments}
This work was supported by the National Natural Science Foundation of China under Grants No.~12675024 and No.~12275193.
\end{acknowledgments}
\end{samepage}
\section*{Data Availability}
The data and code supporting the results of this study are available from the corresponding author upon reasonable request.
\ifdefined\issupplement

\else

\fi
\clearpage
\raggedbottom
\section*{End Matter}
\textit{Correlations between counting intervals.---}Let $X_i=1$ when pair $i$ contains exactly the selected
$(F,A)$ pattern, and $X_i=0$ otherwise. The map $E_h$ associated with this
	event gives the unnormalized final state, whose trace is the pattern probability. The map
$\Phi_{h,\mathrm{pair}}=\e^{2W\mathcal L_h}$ sums all outcomes over the same pair. With
$\mathcal J_FX=\gamma(1-\ell)X_{33}P_4$, $\mathcal J_AX=(\gamma/2)X_{00}P_1$, and
$\mathcal K=\mathcal K_{h,\ell}$, the selected map is
\begin{equation}
	E_h=\int_0^Wdu\int_W^{2W}dv\,
	\e^{(2W-v)\mathcal K}\mathcal J_A
	\e^{(v-u)\mathcal K}\mathcal J_F \e^{u\mathcal K}.
	\label{em:pattern}
\end{equation}
Here $u$ is the time of the $F$ jump within the first half of the pair and $v$ the time of the $A$
	jump within the second half. The selected map $E_h$ and the rejected map
$\Phi_{h,\mathrm{pair}}-E_h$ are completely positive (CP). For $\ell\le W$ and $h^2\le W^{p+2}$, the
selected-event effect obeys $0\le E_h^*(I)\le A_W I$, with $A_W=CW^{p+6}$. This bounds the success
probability for every incoming state. Before pair $i$, the unconditional state is
$\rho_i=\Phi_{h,\mathrm{pair}}^{i-1}P_4$, and the success probability is $p_i=\Tr E_h\rho_i$. The
centered matrix $Y_i=E_h\rho_i-p_i\Phi_{h,\mathrm{pair}}\rho_i$ gives
\begin{align}
	\operatorname{Cov}(X_i,X_j)&=\Tr[E_h\Phi_{h,\mathrm{pair}}^{j-i-1}Y_i],\nonumber\\
	|\operatorname{Cov}(X_i,X_j)|&\le2\sqrt8 A_W^2
	\e^{-2\lambda W(j-i-1)}.\label{em:cov}
\end{align}

The matrix $Y_i$ is traceless, with $\|Y_i\|_1\le2A_W$. Hamiltonian evolution preserves the
Hilbert--Schmidt norm, so the dissipator supplies the same relaxation bound for every $h$. Summing
Eq.~\eqref{em:cov} gives $\operatorname{Var}\sum_iX_i\le Cn(A_W+A_W^2/W)$. The actual initial state $P_4$
shifts the mean from $n\Tr E_h\pi$ by at most $CA_W/W$. Choose $n=\lceil Mh^{-4}W^{p-2}\rceil$ pairs,
where $M$ here is a fixed multiplier. Equation~\eqref{eq:countshift} gives the scaling of the stationary mean and its change due to the link. A midpoint threshold gives each error
at most $C/M$, including the initial-state bound. Observation ends at the deterministic time $2nW$.

\textit{Error bounds at fixed detector settings.---}For the five-bin test, $B_h$ in Eq.~\eqref{eq:fivebin}
describes the selected pattern. Let $\Phi=\e^{5\Delta\mathcal L_h}$ sum all outcomes and $C_h=\Phi-B_h$
sum all rejected patterns. Both retain physical missed jumps; rejection need not mean no registration.
The pattern probabilities under both hypotheses obey $\mu_h<10^{-10}$. Since 50 blocks span a time of ten,
relaxation gives
\begin{equation}
	C_h^{50}\ge_{\rm CP}\left(1-\frac{64}{148}
	-\frac{512\cdot50}{10^{10}}\right)\mathcal D
	>\tfrac12\mathcal D,\label{em:reset}
\end{equation}
Here $\mathcal D(X)=\Tr(X)I/8$, and $\ge_{\rm CP}$ means that the difference between the two sides is completely positive. This
decomposition bounds correlations; it introduces no physical reset. The forward and reversed stationary
processes satisfy the count-comparison assumptions in SM, Sec.~S8. The resulting bound is
$d_{\rm TV}(\mathcal L_\pi(N_M),\operatorname{Pois}(M\mu_h))\le2700\mu_h$ for every finite $M$. The initial
interval of 20 adds at most $b=3/40000$ to this total-variation distance, independently of the subsequent
duration.

For hypothesis $i$, let $\bar\mu_i,r_i$ be the center and radius, both rational, of a rigorous interval for its probability, and
$\mu_i^+$ its upper bound. Use $Q_i=\operatorname{Pois}(M\bar\mu_i)$ as the reference count law. The
initial state, correlations, and probability rounding contribute at most $b$, $2700\mu_i^+$, and
$Mr_i$ to the total-variation distance. Their sum is $\epsilon_i=b+2700\mu_i^++Mr_i$. The minimum
average count error obeys
\begin{equation}
	\left|R_D^{\rm count}-\frac{1-d_{\rm TV}(Q_0,Q_1)}2\right|
	\le\frac{\epsilon_0+\epsilon_1}{2}.\label{em:risk}
\end{equation}
The optimal test for the reference Poisson laws uses one threshold. Applied to the actual count, it
achieves an average error within $(\epsilon_0+\epsilon_1)/2$ of the reference risk. Equation~\eqref{em:risk}
also bounds the best actual count decision from below. Integer Taylor propagation with explicit remainder bounds encloses the pattern probabilities at finite $h$; rational exponential bounds enclose the Poisson tails. Thresholds 25 in C and 11 in S give the attainable upper bounds in
Eq.~\eqref{eq:finite}. The lower bounds cover every decision using the pattern total, but do not optimize
other information in the finite record.
\end{document}


\title{Supplemental Material for ``Detector Placement Relaxes Efficiency Requirements for Detecting a Weak Coherent Link''}
\author{M. H. Yu}
\affiliation{College of Physics and Materials Science, Tianjin Normal University, Tianjin 300387, China}
\author{S. Y. Li}
\affiliation{College of Physics and Materials Science, Tianjin Normal University, Tianjin 300387, China}
\author{X. Z. Zhang}
\email{zhangxz@tjnu.edu.cn}
\affiliation{College of Physics and Materials Science, Tianjin Normal University, Tianjin 300387, China}
\affiliation{Interdisciplinary Center, Tianjin Normal University, Tianjin 300387, China}
\maketitle

\section{Counting experiment and quantities held fixed}
\label{sec:model}
This Supplemental Material contains the derivations used in the Letter.
Sections~\ref{sec:model}--\ref{sec:optimal} give the detected dynamics, short-time
probabilities of background and signal, and the bounds that account for the full return time,
	not only short times. Sections~\ref{sec:absolute-clock}
and~\ref{sec:oct18-joint} derive the observation time when both the missed fraction and the
detector's time-bin width are nonzero. Sections~\ref{sec:finite-example}
and~\ref{sec:oct18-poisson} give the five-bin rule and its finite-time error bounds. All elapsed time is
counted from the single preparation at site 4. Sections~\ref{sec:model}--\ref{sec:optimal} use exact
detection times and directed labels. Section~\ref{sec:absolute-clock} keeps every label and the event
order but replaces each absolute time by its bin index. Section~\ref{sec:count-receiver} retains only the
three unordered class counts per bin. At fixed efficiencies, each record follows by discarding information
from the preceding one. The auxiliary flags and within-bin time offsets used in the proofs are not
experimental data; they are bookkeeping devices available only in the proofs.

\subsection{Model, units, and detection efficiencies}
The sites are $j=0,\ldots,7$, with indices taken modulo eight. Write $P_j=|j\rangle\langle j|$ for the
site projector and $e_j=\{j,j+1\}$ for a ring bond. Set $\hbar=1$ and measure frequencies in units of the
fixed ring hopping $J>0$, with $h=h_{\rm phys}/J$, $\gamma=\gamma_{\rm phys}/J$, and $t=Jt_{\rm phys}$.
The dimensionless Hamiltonian and jump operators are
\begin{align}
	H_h&=H_0+hB,\quad H_0=\sum_j(|j+1\rangle\langle j|+|j\rangle\langle j+1|),\quad B=|0\rangle\langle4|+|4\rangle\langle0|,\label{eq:H}\\
	L_\mu&=\sqrt\gamma M_\mu,\quad M_\mu=|c_\mu\rangle\langle a_\mu|,\qquad \sum_\mu L_\mu^\dagger L_\mu=2\gamma I,\label{eq:jumps}\\
	\mathcal L_hX&=-i[H_h,X]-2\gamma X+\gamma\sum_\mu M_\mu X M_\mu^\dagger.\label{eq:physical}
\end{align}
Here $\mu=(a_\mu\to c_\mu)$ runs over the sixteen directed ring channels, each with dimensionless rate
$\gamma>0$. The link between sites 0 and 4 is purely coherent. The theorem
	of the Letter holds $\gamma$ fixed as $h\downarrow0$; its constants are not uniform as $\gamma$ tends to
zero or infinity.

Each physical jump in channel $\mu$ is missed independently with probability $\delta_\mu$ and registered with efficiency $\eta_\mu=1-\delta_\mu$. Both directions of an edge have the same fraction. C
misses half the jumps in both directions on edges $e_0,e_1$, and S on $e_7,e_0$; every other directed
channel has $\delta_\mu=\ell$. We take $0\le\ell\le\bar\ell\le1/2$, with fixed $\bar\ell>0$. All twelve
protected channels share the missed fraction $\ell$, which is known exactly. The missed fraction in each
high-loss channel remains exactly $1/2$. That fraction does not include an additional $\ell$.

Each experiment tests the null hypothesis $h=0$ against one specified $h>0$. The detector may be chosen to
suit that target, for example by selecting $\ell=\ell(h)$, but its efficiencies are then held fixed under
both hypotheses. The value of $\ell$ is known throughout this test.

\subsection{Detected events and discrimination error}
The unnormalized state between registrations evolves under
\begin{equation}
	\mathcal K_{h,\ell}X=-i[H_h,X]-2\gamma X+\gamma\sum_\mu\delta_\mu M_\mu X
	M_\mu^\dagger.\label{eq:K}
\end{equation}
A missed jump still moves the particle, so its recycling term remains in the conditional generator. Every
physical jump contributes to the anticommutator~\cite{Kiilerich2015}. Thus
$\mathcal K_{h,\ell}=\mathcal L_h-\sum_\mu\mathcal J_\mu$, where
$\mathcal J_\mu(X)=\eta_\mu L_\mu X L_\mu^\dagger$ is the registered-event map. Conditioning on no
registration removes the registered recycling terms without changing physical jump rates. Starting from
$P_b$, $F_{h,\ell}(a,b;t)$ is the unnormalized population at $a$ with no intervening detection. The density
for the next detected event is $w_{h,\ell}$, and the probability of no detection is $S_{h,\ell,b}$. Their
expressions are
\begin{equation}
	F_{h,\ell}(a,b;t)=\langle a|\e^{t\mathcal K_{h,\ell}}(P_b)|a\rangle,\quad
	w_{h,\ell}(\mu,t\mid b)=\gamma\eta_\mu F_{h,\ell}(a_\mu,b;t),\quad
	S_{h,\ell,b}(t)=\Tr \e^{t\mathcal K_{h,\ell}}P_b.\label{eq:F}
\end{equation}
The record identifies the source and destination of each registered jump, and each such jump resets the state to $P_{c_\mu}$.
The operator governing the
	decay of the no-detection probability obeys
\begin{equation}
	\gamma I\le D_{\rm det}=\gamma\sum_\mu\eta_\mu M_\mu^\dagger M_\mu\le2\gamma
	I,\qquad \e^{-2\gamma t}\le S_{h,\ell,b}(t)\le \e^{-\gamma t}.\label{eq:survival}
\end{equation}
For $\rho(t)=\e^{t\mathcal K_{h,\ell}}P_b$,  differentiating the
trace gives $dS_{h,\ell,b}/dt=-\Tr[D_{\rm det}\rho(t)]$. Positivity of $\rho(t)$ then bounds this
derivative between $-2\gamma S$ and $-\gamma S$, giving Eq.~\eqref{eq:survival}. Consequently
$\sum_\mu\int_0^\infty w_{h,\ell}(\mu,t\mid b)dt=1$.

The exact-time reference experiment starts once in $P_4$ and retains every detected direction and
detection time, together with the final interval without a detection. Let $t_i$ be the waiting time
between successive events, with $b_0=4$ and $b_i=c_{\mu_i}$. The probability density for this record is
\begin{equation}
	p_{h,\ell,T}(\mu_1,t_1;\ldots;\mu_n,t_n)=
	\left[\prod_{i=1}^n w_{h,\ell}(\mu_i,t_i\mid b_{i-1})\right]
	S_{h,\ell,b_n}\left(T-\sum_i t_i\right),\qquad \sum_i t_i\le T.\label{eq:likelihood}
\end{equation}
For a record with no detected events, this probability is $S_{h,\ell,4}(T)$. There is no feedback, and the preparation at site 4 is not
	repeated. The optimal equal-prior risk and time are
\begin{equation}
	P_e^*(T,h,\ell)=\tfrac12[1-\TV(\mathbb P_{h,\ell}^{[0,T]},\mathbb P_{0,\ell}^{[0,T]})],\quad
	T_e^D(h,\ell)=\inf\{T:P_e^*(T,h,\ell)\le e\},\quad 0<e<1/2.\label{eq:risk}
\end{equation}
Here $\TV(P,Q)=\tfrac12\int|dP-dQ|$, and $P_e^*$ minimizes the average of the false-alarm probability under
$h=0$ and the missed-link probability under $h>0$. Bounding each error separately gives the same time
order. Their maximum is at most twice their average, so applying the average-error result at target $e/2$
bounds each error by $e$. The dimensionless time $JT_{\rm phys}$ covers the full experiment, including
waits that yield no useful detection.

\subsection{Collection and dynamics with and without the link}
Bidirectional ring jumps cancel on $I$, so $\mathcal L_h(I)=0$ for every 
$h$. The positive-rate ring jumps generate the full
algebra of $8\times8$ matrices. The evolution is therefore primitive, meaning that it
converges to a unique full-rank stationary state. This state is $I/8$, and every
directed stationary flux is $\gamma/8$. The four high-loss directions contribute
a missed flux of $\gamma/4$, and the other twelve a flux of
$3\gamma\ell/2$. Hence
\begin{equation}
	\Phi_{\rm phys}=2\gamma,\qquad \Phi_{\rm missed}=\gamma/4+3\gamma\ell/2,\qquad
	\Phi_{\rm det}/\Phi_{\rm phys}=7/8-3\ell/4.\label{eq:collection}
\end{equation}
At the same $\ell$, both arrangements have the same unconditional generator and collection fraction for
every $h$. Their transient expected counts can differ because the particle starts at site 4.

Let $R|j\rangle=|j-1\rangle$ and $\mathcal R(X)=RXR^\dagger$. At every common value of $\ell$,
\begin{equation}
	\mathcal K_{0,\ell}^{\rm S}=\mathcal R\mathcal K_{0,\ell}^{\rm C}\mathcal
	R^{-1}.\label{eq:rotation}
\end{equation}
Their spectra, including multiplicities, agree. However, $RP_4R^\dagger=P_3$ and
$RBR^\dagger=|7\rangle\langle3|+|3\rangle\langle7|$. The preparation and coherent link remain fixed in the
actual comparison, so this similarity does not make the two labeled record
	distributions identical.

Section~\ref{sec:optimal} shows that the entire distribution of times between detected $3\to4$ returns also
matches. Throughout, $f\asymp g$ and $f=\Theta(g)$ denote two-sided bounds by positive constants; $f=O(g)$
denotes an upper bound. Constants are uniform over the small parameters in each stated domain, at fixed rate and fixed domain limits.

\section{Signal symmetry and short-time transport}
\label{sec:routes}
\subsection{Evenness in the link strength and the leading response}
With $Z=\sum_j(-1)^jP_j$ (which assigns opposite signs to the two sublattices of the ring) and
$\mathcal A(X)=ZX^*Z$, the bipartite ring hopping and the same-sublattice link imply
\begin{equation}
	ZH_0Z=-H_0,\quad ZBZ=B,\quad
	\mathcal A\mathcal K_{h,\ell}\mathcal A=\mathcal K_{-h,\ell}.\label{eq:even-map}
\end{equation}
Every recycling term is invariant under $\mathcal A$, because its sublattice signs
occur twice. The map fixes projectors and population observations, so
$F_{h,\ell}=F_{-h,\ell}$ for every ordered pair and time. Waiting densities, survival
probabilities, and full records are therefore even in $h$. This symmetry relies on real hopping, the
bipartite ring, and the link joining two sites on the same sublattice. It applies to measured
probabilities; a state conditioned on a record can still contain odd powers of $h$. A known perturbation
that breaks the symmetry can allow a linear response, but such a perturbation lies outside the present
comparison.

The link perturbation acts through $\mathcal V(X)=-i[B,X]$. A direct computation gives
\begin{equation}
	\mathcal V^2(P_4)/2=P_0-P_4,\qquad
	\mathcal V^2(P_0)/2=P_4-P_0,\qquad
	\mathcal V^2(P_b)=0\ (b\ne0,4).\label{eq:V2}
\end{equation}
The direct-link amplitude $-iht$ gives leading transferred population $h^2t^2$. Equation~\eqref{eq:V2}
gives this same term through the two Hamiltonian actions in $h^2t^2\mathcal V^2/2$. These actions are
expansion terms, not physical jumps. Joint analyticity on a fixed compact parameter set and the sign
symmetry give the following bounds, uniformly in small $h,t$ and every allowed $\ell$,
\begin{align}
	F_{h,\ell}(0,4;t)-F_{0,\ell}(0,4;t)&=h^2t^2[1-2\gamma t+O(t^2)],\label{eq:response}\\
	|F_{h,\ell}(a,b;t)-F_{0,\ell}(a,b;t)|&\le Ch^2t^3
	\quad(a\ne b,\ \{a,b\}\ne\{0,4\}).\label{eq:other-response}
\end{align}

The factor $-2\gamma t$ in Eq.~\eqref{eq:response} comes from the common damping
$\e^{-2\gamma t}$ between physical jumps; neither a
three-step Hamiltonian path nor a single missed-jump contribution changes the diagonal
coefficient at this order. The reverse endpoint, with the roles of 0 and 4
exchanged, has the same coefficient. The diagonal pairs $(0,0)$ and $(4,4)$ have an $h^2t^2$ coefficient of $-1$, but their null populations are bounded below near zero.
These diagonal entries are not part of the off-diagonal bound. Higher
powers of $h$ start at higher time order; there is no $\ell$-dependent
or uncontrolled correction to the exact $h^2t^2$ coefficient.

\subsection{Route coefficients and a uniform analytic remainder}
Between physical jumps, the propagator is $U_0(t)=\e^{-\gamma t}\e^{-iH_0t}$. Expanding Eq.~\eqref{eq:K} order by order in the number of missed
	jumps gives positive contributions with rates $\gamma\delta_\mu$. Within each interval between such jumps,
coherent amplitudes must be added before the squared modulus is taken. Contributions from different
missed-jump histories are then added as probabilities. A unique shortest coherent path of $d$ bonds has
leading amplitude $(-i)^dt^d/d!$ and population $t^{2d}/(d!)^2$. When several shortest paths exist, their
amplitudes add. In particular, the two four-bond arcs from 4 to 0 give $4t^8/(4!)^2=t^8/144$. The coherent
segments in the leading polynomials below have at most three bonds and are unique.

We now calculate coefficients as well as time powers. A history consists of missed jumps and the coherent
segments between them. For $r$ missed jumps and segment lengths $d_0,\ldots,d_r$, integration over the
ordered jump times gives
\begin{equation}
	\frac{\prod_{q=0}^r (2d_q)!/(d_q!)^2}{(r+2\sum_qd_q)!}
	\prod_{\text{hidden edges}}\gamma\delta_j,\label{eq:simplex}
\end{equation}
Here the product over hidden edges means the rates of the missed jumps in that history, that is, over
	the edges crossed by a missed jump. To obtain the coefficient, denote the coherent durations by
$u_0,\ldots,u_r\ge0$, with $\sum_q u_q=t$. Their integral is
\begin{equation*}
	\int_{\sum_q u_q=t}\prod_{q=0}^r\frac{u_q^{2d_q}}{(d_q!)^2}\,du_0\cdots du_{r-1}
	=\frac{t^{r+2\sum_qd_q}}{(r+2\sum_qd_q)!}\prod_{q=0}^r\frac{(2d_q)!}{(d_q!)^2}.
\end{equation*}

Multiplication by the missed-jump rates gives Eq.~\eqref{eq:simplex}. Positive real hopping gives all shortest
coherent walks the same phase. The short-time path construction of Ref.~\cite{Szigeti2019} applies with the
site resets caused by missed jumps.

For C, the arc $4\to3\to2\to1\to0$ crosses the two high-loss bonds by missed jumps. The other two bonds may
be crossed by two missed jumps, by one missed jump and one coherent segment, or entirely coherently.
Equation~\eqref{eq:simplex} gives the contributions below. In the joint expansion, each factor of $t$ and
each factor of $\ell$ counts as one degree. The opposite arc contains only protected bonds and has higher
joint degree. For S, each four-bond arc contains one high-loss bond; summing the two arcs gives
\begin{align}
	Q_{\rm C}(t,\ell)&=\frac{\gamma^4}{96}\ell^2t^4+\frac{\gamma^3}{120}\ell
	t^5+\frac{\gamma^2}{480}t^6,\label{eq:QC}\\
	Q_{\rm S}(t,\ell)&=\frac{\gamma^4}{24}\ell^3t^4+\frac{\gamma^3}{20}\ell^2t^5+
	\frac{\gamma^2}{45}\ell t^6+\frac{\gamma}{252}t^7.\label{eq:QS}
\end{align}
At $\ell=0$, the C route has coherent run lengths $(2,0,0)$
and two missed-jump rates $\gamma/2$, giving $6(\gamma/2)^2/6!=\gamma^2/480$. Each S arc has
run lengths $(3,0)$ or $(0,3)$ and one missed-jump rate $\gamma/2$; the two arcs give
$2\times20(\gamma/2)/7!=\gamma/252$. For the S term with two protected bonds crossed
coherently and one crossed by a missed jump, the factors from the coherent segments are $6,4,6$, the
	products $\prod_q(2d_q)!/(d_q!)^2$ for the three admissible splittings of the run lengths. Their sum,
divided by $6!$, produces $1/45$; replacing the runs by independent single-edge coherent probabilities would
be incorrect. Both endpoint orientations have these coefficients.

With this degree convention, the leading terms have degree $6$ for C and $7$ for S. For every $\ell$, the
null endpoint population starts at order $t^4$ or higher, since a route confined to the ring crosses at
least four bonds. After factoring $t^4$, no monomial in the remaining analytic function has degree below
$p$. Subtracting the leading homogeneous polynomial leaves a remainder bounded by $Ct^4(t+\ell)^{p+1}$.
Because all coefficients of $Q_D$ are positive,
	$Q_D\ge c\,t^4(\ell+t)^p$ for a constant $c>0$, and therefore
\begin{equation}
	F_{0,\ell}^D(0,4;t)=Q_D(t,\ell)[1+O(t+\ell)],\qquad
	F_{0,\ell}^D\asymp t^4(\ell+t)^p,\quad p=2\ (\mathrm C),\ 3\ (\mathrm S).\label{eq:joint}
\end{equation}
The exact coefficients of order $t^4$ at fixed $\ell$ are
\begin{equation}
	[t^4]F_{0,\ell}^{\rm C}=\gamma^4(\ell^2/96+\ell^4/24),\qquad
	[t^4]F_{0,\ell}^{\rm S}=\gamma^4\ell^3/24.\label{eq:t4}
\end{equation}
The additional C term $\gamma^4\ell^4t^4/24$ comes from four missed jumps on the fully protected arc. It has
joint degree eight and therefore lies beyond the degree-six polynomial $Q_{\rm C}$. Thus the joint
$O(t+\ell)$ remainder is essential. The polynomials in
Eqs.~\eqref{eq:QC} and~\eqref{eq:QS} are leading terms in the joint limit. They are
neither exact binomial powers nor, at fixed $\ell$, Taylor polynomials
with relative $O(t)$ error.

\section{Uniform information bounds for every registered wait}
\label{sec:hellinger}
\subsection{Short waiting times and a binary detection rule}
We quantify the difference between waiting laws with the squared Hellinger distance
$\HH^2(P,Q)=\int(\sqrt{dP}-\sqrt{dQ})^2$ and affinity $\aff(P,Q)=1-\HH^2/2$. This convention omits the factor $1/2$ sometimes included in the definition of squared Hellinger distance; all affinity and total-variation bounds below use the convention stated here. The target
next-event density from $P_4$ is $f_{h,\ell}(t)=(\gamma/2)F_{h,\ell}(0,4;t)$ for
both arrangements. On a sufficiently short fixed interval
$(0,\tau)$, the increment of $f_{h,\ell}$ over
	$f_{0,\ell}$ is comparable to $h^2t^2$. Rationalizing the difference of square roots
and using Eq.~\eqref{eq:joint} gives
\begin{equation}
	\int_0^\tau(\sqrt{f_{h,\ell}}-\sqrt{f_{0,\ell}})^2dt
	\asymp h^4\int_0^\tau\frac{t^2dt}{t^2(\ell+t)^p+h^2}.\label{eq:singular-integral}
\end{equation}
For completeness, set $u_j=\int_0^\tau f_{j,\ell}(t)dt$, $j=0,h$. Merging every other
outcome into one atom gives the normalized squared distance
\begin{equation*}
	\HH^2_{\rm end}=\int_0^\tau(\sqrt{f_{h,\ell}}-\sqrt{f_{0,\ell}})^2dt
	+(\sqrt{1-u_h}-\sqrt{1-u_0})^2.
\end{equation*}
Choose $\tau$ small enough that $u_j\le1/2$ uniformly. Since $u_h-u_0=O(h^2)$, the
final term is $O(h^4)$ and does not change the order of the
	endpoint information.

Set $\rho=h^{2/(p+2)}$, $L=\ell+\rho$, and $R_D=h^4/L^{p-1}$. Here $\rho$ is the signal--background
crossover time at zero missed fraction, $L$ combines the loss scale
	$\ell$ with $\rho$, and $R_D$ is the resulting information order. For $2L\le\tau$, integration over
$[L,2L]$ gives a lower bound $cR_D$, since $h^2\le L^{p+2}$. The pointwise upper bound is
\begin{equation}
	C\min\{h^2t^2,h^4/(\ell+t)^p\}.\label{eq:endpoint-upper}
\end{equation}
If $\ell\ge\rho$, integrate the second entry of the
minimum over $(0,\tau)$ to obtain $Ch^4\ell^{1-p}$. If $\ell\le\rho$, split at $\rho$;
both parts are $O(h^4\rho^{1-p})$, because $h^2\rho^3=h^4\rho^{1-p}$. Hence the endpoint
information is uniformly $\Theta(R_D)$ without a leading logarithm. The rate-dependent
coefficients change the multiplicative constants but leave these joint
scalings unchanged.

A normalized binary observation attains the same order. Record success if the next registered event is
$0\to1$ and occurs before $\tau_{\rm bin}=b_{\rm bin}L\le\tau$, with fixed positive $b_{\rm bin}$. Every other
outcome is a failure. For success probabilities $s_0,s_h$, Eqs.~\eqref{eq:joint} and~\eqref{eq:response} give
\begin{equation}
	s_0\asymp L^{p+5},\qquad s_h-s_0\asymp h^2L^3,\qquad s_h+s_0\asymp L^{p+5}.\label{eq:binary-scale}
\end{equation}
The last comparison uses $h^2\le L^{p+2}$. Rationalizing the square-root difference for
the success outcome yields
\begin{equation}
	\HH^2(\operatorname{Bern}(s_h),\operatorname{Bern}(s_0))
	\ge c\frac{h^4L^6}{L^{p+5}}=cR_D.\label{eq:binary}
\end{equation}
Including the failure outcome normalizes the binary probability law. Its Hellinger upper
bound follows by the data-processing inequality from the
complete-wait bound below. The receiver accumulates these binary marks in a likelihood
test; a single rare mark need not determine the decision.

\subsection{All other reset/source pairs}
A necessary time bound must control every next-event channel, not only the target. We therefore compare
waiting laws for every reset/source pair. Summing positive trajectory contributions gives the uniform
lower bound
\begin{equation}
	F_{0,\ell}(a,b;t)\ge ct^4(\ell+t)^3\quad(0<t\le\tau,\ 0\le\ell\le\ell_0)\label{eq:allpairs}
\end{equation}
for every ordered pair and either arrangement. To see this, at distances $d\le3$, choose a shortest path and
sum over the two options, a coherent
	crossing or a missed jump, on each edge. Their nonnegative contributions are of scale at least
$t^d(\ell+t)^d$, which dominates the displayed bound when $t,\ell+t\le1$. The missed-jump rate on a high-loss
bond is bounded below by a fixed multiple of that on a protected bond. At distance four, one of the two arcs
contains a high-loss edge. Fix that crossing as a missed jump on that high-loss bond and sum the two choices
on its other three edges, giving $t^4(\ell+t)^3$. Diagonal zero-jump histories are bounded away from zero.
There are finitely many route families, allowing one common positive constant.

For off-diagonal pairs other than the two link orientations, Eqs.~\eqref{eq:other-response}
and~\eqref{eq:allpairs} imply the Hellinger-density upper bound
\begin{equation}
	C\min\{h^2t^3,h^4t^2/(\ell+t)^3\}.\label{eq:regular-density}
\end{equation}
If $\ell\ge\sqrt h$, the second entry of the minimum integrates over $(0,\tau)$ to $O(h^4[1+\log(1/\ell)])$.
Otherwise split at $\sqrt h$ to obtain $O(h^4[1+\log(1/\sqrt h)])$. Thus all such pairs contribute at most
\begin{equation}
	Ch^4[1+\log(1/(\ell+\sqrt h))].\label{eq:coarse-log}
\end{equation}
This upper bound suffices for the comparison; it does not establish a logarithmic contribution. Both endpoint
orders dominate it: with $L_{\rm C}=\ell+\sqrt h$ and $L_{\rm S}=\ell+h^{2/5}$,
\begin{equation}
	1+\log(1/L_{\rm C})\le C/L_{\rm C},\qquad
	L_{\rm S}^2[1+\log(1/L_{\rm C})]\le C.\label{eq:log-dominated}
\end{equation}
For the second inequality, $L_{\rm S}^2\le C L_{\rm C}$ for small $h$ and $x[1+\log(1/x)]$ is bounded.
Diagonal pairs contribute $O(h^4)$; the two link orientations obey their endpoint bounds.

\subsection{Positivity improvement at every positive time}
To establish positivity at every $t>0$, use the common high-loss bond $0\leftrightarrow1$, whose
missed-jump rate is $\gamma/2$ in each direction. The coherent evolutions from sites 0 and 1 together span
the full site space. Such a pair is called jointly cyclic. Writing $v_j=|j\rangle$, this follows from
\begin{align}
	v_2&=H_hv_1-v_0,&v_3&=H_hv_2-v_1,&v_4&=H_hv_3-v_2,\nonumber\\
	v_5&=H_hv_4-v_3-hv_0,&v_6&=H_hv_5-v_4,&v_7&=H_hv_6-v_5.\label{eq:cyclicity}
\end{align}
These relations generate every site vector by repeated application of $H_h$ to $v_0$ and $v_1$. Thus a state orthogonal to every coherent evolution of both sites must vanish.
Hermiticity gives the corresponding statement for observations at those two sites.

Fix nonzero vectors $\xi,v$ and $t>0$, and put $U_h(s)=\e^{-iH_hs}$. At least one source
overlap $f_i(s)=\langle i|U_h(s)|\xi\rangle$, $i=0,1$, is a nonzero analytic function.
Suppose it is $f_0$. If $g_1(r)=\langle v|U_h(r)|1\rangle$ is not identically zero, the
one-missed-jump history $0\to1$ has a positive contribution proportional to
\begin{equation}
	\int_0^t |g_1(t-s)f_0(s)|^2ds>0.\label{eq:one-hidden}
\end{equation}
If $g_1$ vanishes identically, $g_0$ does not. The two-missed-jump history $0\to1\to0$
contributes
\begin{equation}
	\int_{0<s_1<s_2<t}|g_0(t-s_2)\langle1|U_h(s_2-s_1)|1\rangle
	f_0(s_1)|^2ds_1ds_2>0.\label{eq:two-hidden}
\end{equation}
Choose $s_1$ away from the isolated zeros of $f_0(s_1)g_0(t-s_1)$ and $s_2$ just above
$s_1$; the middle factor approaches unity, leaving a nonzero open subset of
the simplex. For a nonzero $f_1$, exchange the two sites. Positive scalar damping and missed-jump
rates do not affect the argument. Spectral decomposition of a positive input therefore gives
\begin{equation}
	\e^{t\mathcal K_{h,\ell}}(X)\succ0\quad\text{for every }X\succeq0,\ X\ne0,\ t>0.\label{eq:positivity}
\end{equation}
Adding the other missed-jump recycling terms preserves positivity. This applies to all
$h$, and also establishes the fixed-parameter tails needed later.

\subsection{Long waits and the remaining missed-fraction interval}
Positive-map Perron theory~\cite{Evans1978} gives full-rank positive right and left
matrices $R_{h,\ell},L_{h,\ell}$, a simple dominant exponent $-\kappa_{h,\ell}$, and a strict gap between the dominant and subleading real parts. Normalize $\Tr(LR)=1$. For
fixed parameters,
\begin{equation}
	\e^{t\mathcal K_{h,\ell}}(X)=\e^{-\kappa_{h,\ell}t}
	[R_{h,\ell}\Tr(L_{h,\ell}X)+O(\e^{-d_{h,\ell}t})],\quad \gamma\le\kappa_{h,\ell}<2\gamma.\label{eq:Perron}
\end{equation}
Polynomial factors from subleading modes are absorbed by choosing a slightly smaller decay gap. Tracing the
Perron equation gives $\kappa_{h,\ell}<2\gamma$: recycling on the common high-loss bond subtracts a strictly
positive term from $2\gamma\Tr R_{h,\ell}$.

Use the null decay exponent at $\ell=0$,
$\lambda_*=\kappa_{0,0}$, for which $0<\lambda_*<2\gamma$. Full rank gives
$F_{0,0}(a,b;t)\ge ce^{-\lambda_*t}$ for all pairs and sufficiently
large $t$; on compact intervals bounded away from $t=0$ the function has a positive
minimum. The $\ell$-dependent recycling is a positive
perturbation, so its Dyson expansion implies
$\e^{t\mathcal K_{0,\ell}}X\succeq \e^{t\mathcal K_{0,0}}X$ for $X\succeq0$. These lower
bounds on the Hellinger denominator hold uniformly for every allowed $\ell$.

Equation~\eqref{eq:survival} and the Jordan decomposition give Hermitian trace-norm
contraction
\begin{equation}
	\|\e^{t\mathcal K_{h,\ell}}X\|_1\le \e^{-\gamma t}\|X\|_1,\qquad X=X^\dagger.\label{eq:contraction}
\end{equation}
Since $\|\mathcal VX\|_1\le2\|X\|_1$, two insertions of $\mathcal V$ in Duhamel's formula bound the second
derivative in $h$ by $4t^2\e^{-\gamma t}$, at fixed detector efficiencies. The first derivative of each
measured population vanishes at $h=0$ by sign symmetry. Taylor's integral remainder gives
\begin{equation}
	|F_{h,\ell}-F_{0,\ell}|\le2h^2t^2\e^{-\gamma t}.\label{eq:Duhamel}
\end{equation}
For a fixed $t_0>0$ independent of $h$ and $\ell$, the tail integral
$\sum_\mu\int_{t_0}^\infty(w_{h,\ell}-w_{0,\ell})^2/w_{0,\ell}\,dt$ is an upper bound on the Hellinger tail.
This chi-square upper bound is at most
\begin{equation}
	Ch^4\int_{t_0}^\infty t^4\e^{-(2\gamma-\lambda_*)t}dt=O(h^4),\label{eq:tail}
\end{equation}
uniformly in $\ell$. The same bound holds on compact time intervals
away from zero.

For $\ell$ in a fixed interval $\ell_0\le\ell\le\bar\ell$,
all missed-jump rates are bounded below. Paths consisting entirely of missed jumps give
$F_{0,\ell}(a,b;t)\ge ct^{d(a,b)}\ge ct^4$ on a short interval. The uniform
$O(h^2t^2)$ response makes its Hellinger integral $O(h^4)$. Conversely a fixed
sufficiently short target bin has a success increment of at least $ch^2$ and a
uniformly bounded denominator, hence a binary Hellinger cost of at least $ch^4$.
For both arrangements $R_D$ is comparable to $h^4$ on this
compact interval. Combining the small-$\ell$ analysis with the bound
for all positive times proves
\begin{equation}
	\max_b\HH^2(w_{h,\ell}(\cdot,\cdot\mid b),w_{0,\ell}(\cdot,\cdot\mid b))\le CR_D(h,\ell),\quad
	R_{\rm C}=\frac{h^4}{\ell+\sqrt h},\quad R_{\rm S}=\frac{h^4}{(\ell+h^{2/5})^2}.\label{eq:one-wait-bound}
\end{equation}
A normalized binary mark has a matching lower bound from $P_4$ over the whole interval.
All constants are independent of $h,\ell$ in the stated domain.

\section{Minimum observation time including physical returns}
\label{sec:optimal}
\subsection{Lower bound including the last interval without detection}
For any fixed observed history, the last registered destination determines the same reset state under both
hypotheses. Equation~\eqref{eq:one-wait-bound} therefore bounds each conditional affinity below by
$1-CR_D/2$. Successive integration over histories gives
\begin{equation}
	\aff(\mathbb P_{h,\ell}^{(n)},\mathbb P_{0,\ell}^{(n)})\ge(1-CR_D/2)^n,\qquad
	\HH^2(\mathbb P_{h,\ell}^{(n)},\mathbb P_{0,\ell}^{(n)})\le CnR_D.\label{eq:composition}
\end{equation}
At each step the same conditional bound applies, giving the first inequality by induction. The inequality
$1-(1-x)^n\le nx$ gives the second. This includes the initial preparation $P_4$; independence of successive
waits is not required.

Since the physical jump intensity is $2\gamma$, detected counts are dominated by
$\operatorname{Pois}(2\gamma T)$. Set $n=\lceil4\gamma T+\log(1/h)\rceil+1$. Unless
the detected count exceeds this cutoff, these complete waits determine the record
through $T$, including the wait that crosses $T$ and the part of that wait before the
endpoint. They form an auxiliary, more informative record for the converse; the
physical receiver never observes beyond $T$. Replacing all records beyond the cutoff by
one common symbol and applying data processing with $\TV\le\HH$ gives
\begin{equation}
	\TV(\mathbb P_{h,\ell}^{[0,T]},\mathbb P_{0,\ell}^{[0,T]})\le\sqrt{CnR_D}
	+2\Pr\{\operatorname{Pois}(2\gamma T)\ge n\}.\label{eq:record-converse}
\end{equation}
The factor two accounts for one discarded-tail probability under each hypothesis. The final no-event interval
is included through its survival probability in Eq.~\eqref{eq:likelihood}. Since $R_D\log(1/h)\to0$
uniformly as $h\downarrow0$, at $T=c/R_D$ a sufficiently small fixed $c>0$ makes the right-hand side smaller
than $1-2e$; the Poisson term vanishes uniformly as $h\downarrow0$. Thus $T_e^D\ge c_e/R_D$ for every
fixed error target. The bound applies to every complete-record decision rule, including those using initial
waits, additional labels, correlations, and censoring at the final time.

\subsection{Exact return-time distribution and moments}
Registered arrivals $3\to4$ reset the state to $P_4$ in both
arrangements. The selected-event map and the generator conditioned on no such event are
\begin{equation}
	\mathcal J_{\rm ret}(X)=\gamma(1-\ell)X_{33}P_4,\qquad
	\mathcal Q_{h,\ell}=\mathcal L_h-\mathcal J_{\rm ret}.\label{eq:return-generator}
\end{equation}
All other physical events, whether observed or missed, are recycled. This generator is
identical for C and S, so the full return-time distributions
agree at each common $(h,\ell)$:
\begin{equation}
	\Pr(\tau>t)=\Tr \e^{t\mathcal Q_{h,\ell}}P_4,\qquad
	f_\tau(t)=\gamma(1-\ell)\langle3|\e^{t\mathcal Q_{h,\ell}}P_4|3\rangle.\label{eq:return-law}
\end{equation}
The distribution can depend on $h$, even though its mean does not. The generator conditioned on no registered
return is positivity improving by the same argument using missed jumps on the common bond. Tracing the Perron
equation for its positive-definite eigenmatrix gives a strictly negative spectral bound, because the
selected-event loss is $\gamma(1-\ell)R_{33}>0$. Its resolvent therefore exists. Using $\mathcal L_h(I)=0$ gives
\begin{equation}
	\mathcal Q_{h,\ell}(I)=-\gamma(1-\ell)P_4,\qquad
	\int_0^\infty \e^{t\mathcal Q_{h,\ell}}P_4dt=-\mathcal Q_{h,\ell}^{-1}P_4=\frac{I}{\gamma(1-\ell)}.\label{eq:return-resolvent}
\end{equation}
Taking the trace and the first time moment yields
\begin{equation}
	\mathbb E\tau=\frac8{\gamma(1-\ell)},\qquad
	\mathbb E\tau^2=2\Tr\mathcal Q_{h,\ell}^{-2}P_4.\label{eq:return-moments}
\end{equation}
All intervening evolution and missed returns are counted. Using $8/\gamma$ at nonzero $\ell$ would ignore the returns that are never
	registered.

\subsection{A test using the actual return process}
Each detected return places the particle in $P_4$. Rank-one resetting and the strong Markov property
therefore make the complete cycles between returns independent and identically distributed under each
hypothesis. The initial preparation is the same state. In each cycle, retain the binary outcome of its first
detected event, using Eq.~\eqref{eq:binary} at small $\ell$ or the fixed short-bin rule on the compact remainder of the $\ell$ domain. If that event is itself the terminal return,
the binary outcome is failure. A cycle includes every later event and all elapsed time until the next
registered return. Its binary outcome may be correlated with its duration.

The binary one-cycle affinity is at most $1-cR_D/2$. With $n=\lceil A/R_D\rceil$, the
equal-prior product-likelihood error is at most $\frac12\e^{-cA/2}$.

Uniformly in $\ell$, $\mathbb E\tau\le M=8/[\gamma(1-\bar\ell)]$. At $T=Bn$, Markov's inequality gives
$\Pr(S_n>T)\le M/B$ under both hypotheses. Choose $A$ so that the unrestricted $n$-cycle test has error below
$e/2$, and $B$ so that failure to complete $n$ cycles by $T$ has probability below $e/2$. At $T$, use the
first $n$ binary outcomes if available; otherwise make a fixed fallback decision. Comparison with the
unrestricted test bounds the error without conditioning its product law on timely completion. Unsuccessful
cycles, missed returns, and the final unfinished cycle are all counted, proving $T_e^D\le C_e/R_D$.

\begin{theorem}[Exact-time counting record] Fix $\gamma>0$, $0<\bar\ell\le1/2$, and $0<e<1/2$. There exists
	$h_0>0$ such that, uniformly for $0<h\le h_0$ and $0\le\ell\le\bar\ell$, the optimum
	over all rules using the specified complete registered record satisfies
	\begin{equation}
		T_e^{\rm C}(h,\ell)=\Theta\!\left(\frac{\ell+\sqrt h}{h^4}\right),\qquad
		T_e^{\rm S}(h,\ell)=\Theta\!\left(\frac{(\ell+h^{2/5})^2}{h^4}\right).\label{eq:optimal-time}
	\end{equation}
\end{theorem}
The constants depend on the fixed rate, the bound $\bar\ell$, and
target error, not on $h$ or $\ell$. The optimization is over decision rules
for the exact-time, directed-label counting record and known signal alternative defined in
Sec.~\ref{sec:model}. The observation retains all detected jumps and their exact times, rather than just their
total number. The uniform
estimates determine the time scalings; optimal prefactors and
finite-time risk thresholds require further analysis.

\section{Time-resolution bound at zero missed fraction}
\label{sec:absolute-clock}
Set $\ell=0$ and fix $\gamma>0$. To prove a necessary time bound, temporarily give the observer more data:
every directed label and the order of events within each absolute clock bin. A lower bound for this richer
record also applies to the three-count record in Sec.~\ref{sec:count-receiver}. The derivation covers every
reset site and an experiment ending partway through its last bin. Only the protected channels become
perfectly detected; the four high-loss channels still miss half their jumps.

\subsection{Detector record and information bound}
Sites are $0,\ldots,7$ modulo eight, $P_j=|j\rangle\langle j|$, and time is in units of $J^{-1}$. Throughout
this section, $\mathcal K_h=\mathcal K_{h,0}$. The model is
\begin{align}
	H_h&=\sum_{j=0}^{7}(|j+1\rangle\langle j|+\mathrm{H.c.})
	+h(|0\rangle\langle4|+|4\rangle\langle0|),\nonumber\\
	L_{a\to c}&=\sqrt\gamma\,|c\rangle\langle a|,\quad c=a\pm1,\nonumber\\
	\mathcal K_hX&=-i[H_h,X]-2\gamma X
	+\gamma\sum_{a\to c}\delta_{ac}|c\rangle\langle a|X|a\rangle\langle c|.
	\label{clock:eq:model}
\end{align}
At $\ell=0$, C has missed fraction $\delta=1/2$ in
both directions on bonds $\{0,1\}$ and $\{1,2\}$; S has it on $\{7,0\}$ and
$\{0,1\}$. All other bonds are perfectly detected. Set $\eta_\mu=1-\delta_\mu$. Starting
from the state $P_b$ left by a detected jump,
\begin{equation}
	q_\theta(\mu,t\mid b)=\gamma\eta_\mu
	\langle a_\mu|\e^{t\mathcal K_h}P_b|a_\mu\rangle,
	\qquad \theta=h^2.
	\label{clock:eq:q}
\end{equation}

The directed label $\mu=a_\mu\to c_\mu$ fixes the state $P_{c_\mu}$ after the next
detected jump. The initial state is $P_4$. Let $s_i=\sum_{j=1}^i t_j$ be the absolute time
of the $i$th detection, where $t_j$ is an inter-event waiting time as in
Eq.~\eqref{eq:likelihood}. Up to the deterministic
time $T$, the observation retains every ordered label and
$B_i=\lfloor s_i/\Delta\rfloor$ for $s_i<T$, without the within-bin time offsets. Its law is
$P_\theta^{D,T,\Delta}$. Write $T_{e,D}^{\rm ord}(h,\Delta)$ for the minimum deterministic horizon at which this
ordered-label binned record permits equal-prior error at most $e$.

For $D=\mathrm C,\mathrm S$, respectively, define
\begin{equation}
	(s_D,a_D,\nu_D)=(4,1,7/4),\ (5,2,8/5),\qquad
	R_D(\theta,\Delta)=\frac{\theta^2}{(\Delta+\theta^{1/s_D})^{a_D}}.
	\label{clock:eq:rate}
\end{equation}
The exponent $s_D$ sets the crossover time $\theta^{1/s_D}$. The exponent $a_D$ controls information loss from
coarse timing, and $\nu_D$ gives the power of $\theta$ in the
	fine-clock information.
\begin{theorem}[Information and observation-time bounds]\label{clock:thm:frontier}
	Fix $\gamma>0$. There exist $C,\theta_0,\Delta_0>0$ such that, uniformly for $T\ge0$,
	$0\le\theta\le\theta_0$, and $0<\Delta\le\Delta_0$,
	\begin{equation}
		D_{\mathrm{KL}}(P_\theta^{D,T,\Delta}\Vert P_0^{D,T,\Delta})
		+D_{\mathrm{KL}}(P_0^{D,T,\Delta}\Vert P_\theta^{D,T,\Delta})
		\le C(T+1)R_D(\theta,\Delta)+C\theta^{\nu_D}.
		\label{clock:eq:main}
	\end{equation}
	At full-bin horizons $T=M\Delta$, with integer $M\ge0$, the additive term can be omitted. For every fixed
	equal-prior error $0<e<1/2$, the minimum observation time for the known pair $0$
	versus $h>0$, over decision rules that fix the
		time horizon in advance, satisfies
	$T_{e,D}^{\rm ord}(h,\Delta)\asymp R_D(h^2,\Delta)^{-1}$ uniformly in the stated parameter range.
\end{theorem}
The information bound holds over the stated parameter range and leads to a minimum
observation time at fixed error. It does not require a limiting relative-entropy rate.
Throughout this section, the protected bonds are perfectly detected and $\gamma$
is fixed.

\subsection{Short-time probabilities for every reset site}
The short-time probabilities for every reset site follow from the exact generator
recurrence below and the positive-time estimates in Sec.~\ref{sec:hellinger}. The
transformation $X\mapsto ZX^*Z$, $Z=\sum_j(-1)^jP_j$ (which assigns opposite signs to the two
	sublattices), relates $\mathcal K_h$ and $\mathcal K_{-h}$ while preserving each input projector and
observed population. Each
scalar probability is therefore analytic in $\theta$ near zero.

A row below means a fixed reset site and source site for the next detected jump. Let $m$ be the null
population's leading time power and $r$ the first time power in its $\theta$ coefficient. We call a row
singular when $r<m$, so the signal correction starts at a lower power than the background.
Table~\ref{clock:tab:rows} lists all such pairs; each unordered pair includes both orientations.
Coefficients $c_0,c_1$ refer to populations before multiplication by $\gamma\eta_\mu$:

\begin{table}[!htbp]
	\centering
	\caption{All reset/source pairs whose signal term starts at a lower time order than the null term, at
		$\ell=0$. Coefficients refer to dimensionless populations; each unordered pair includes both
		orientations.}
	\label{clock:tab:rows}
	\begin{tabular}{cclll}
		\toprule
		Arrangement & $(m,r)$ & Unordered reset/source pairs & $c_0$ & $c_1$\\
		\midrule
		C & $(6,2)$ & $\{0,4\}$ & $\gamma^2/480$ & $1$\\
		C & $(6,4)$ & $\{0,5\},\{4,7\}$ & $1/36$ & $1/4$\\
		C & $(5,3)$ & $\{1,4\}$ & $\gamma/40$ & $\gamma/6$\\
		C & $(7,5)$ & $\{1,5\}$ & $\gamma/252$ & $\gamma/40$\\
		\midrule
		S & $(7,2)$ & $\{0,4\}$ & $\gamma/252$ & $1$\\
		S & $(6,3)$ & $\{1,4\},\{4,7\}$ & $1/36$ & $\gamma/6$\\
		S & $(5,4)$ & $\{0,3\},\{0,5\}$ & $\gamma/40$ & $1/4$\\
		S & $(6,5)$ & $\{1,5\},\{3,7\}$ & $\gamma^2/480$ & $\gamma/40$\\
		\bottomrule
	\end{tabular}
\end{table}

For these rows, the exact local form is
\begin{equation}
	F_h(a,b;t)=c_0 t^m A(t)+c_1\theta t^r B(\theta,t),\qquad
	A(t)=1+O(t),\quad B(\theta,t)=1+O(t),
	\label{clock:eq:factorraw}
\end{equation}
with positive $c_0,c_1$ and uniform analytic remainders. All other pairs have $m\le r$
and $F_h(a,b;t)=t^m G(\theta,t)$ with $G$ analytic and bounded above and below by
positive constants. This includes diagonal rows with a negative leading signal
coefficient.

The short-time orders follow directly from the generator recurrence. Let
$T_n=(2\mathcal K_h)^nP_b$. Its entries are polynomials in $h,\gamma$ with coefficients in the Gaussian integers $\mathbb Z[i]$, and
\begin{equation}
	F_h(a,b;t)=\sum_{n\ge0}\frac{t^n}{2^n n!}(T_n)_{aa},\qquad
	T_{n+1}=2\mathcal K_hT_n.\label{clock:eq:recurrence}
\end{equation}
The recurrence consists of the left and right terms of the ring commutator, with coefficients $\mp2i$, the link commutator with $\mp2ih$, damping $-4\gamma$, and, on a
high-loss directed edge only, recycling $\gamma E_{cc,aa}$. Each polynomial has total degree at most
$n$. Exact iteration through $n=9$ for all eight inputs and both
arrangements establishes:
\begin{enumerate}
	\item the first null coefficient of every pair is positive, with $0\le m\le7$;
	\item for a singular row, every positive power of $\theta$ vanishes at time orders below
	$r$, and at order $r$ the only such term is exactly $c_1\theta$ from the table;
	\item for every remaining row, every positive power of $\theta$ vanishes below $m$.
\end{enumerate}

At time order $n$, the degree in $h$ is at most $n$. The finite recurrence therefore checks every possible
positive power of $h$ at the orders required for Eq.~\eqref{clock:eq:factorraw}. Joint analyticity supplies the
uniform remainder and derivative bounds. Higher link insertions and interference are included, rather than
discarded by a shortest-path assumption. The table is reproducible from Eq.~\eqref{clock:eq:recurrence}:
\texttt{check\_row\_orders.py} uses integer and rational arithmetic, and \texttt{row\_orders.json} records
every pair's leading orders and coefficients. These data and code are available from the corresponding author
upon reasonable request.

For a singular row, put $s=m-r$. Let $u$ denote the variable value of $\theta=h^2$. The score of a density or
mass $q_u$ means its logarithmic derivative $\partial_u\log q_u$. There are positive constants $A,B$ such that
\begin{align}
	q_u(t)&=t^r(A t^s+B u)\e^{\psi_u(t)},
	&|\partial_t\psi_u(t)|&\le C,\nonumber\\
	|\partial_u\log q_u(t)|&\le C\{1+t+(t^s+u)^{-1}\},
	&q_u(t)&\le C(t^m+u t^r)\quad(t\text{ small}).
	\label{clock:eq:rowbounds}
\end{align}
Regular rows have $q_u(t)=t^m \e^{\psi_u(t)}$, $0\le m\le7$, whose logarithmic derivative in $u$ is bounded by
$C(1+t)$. The constants are uniform
over the finite set of reset/source pairs. Only the exceptional $(m,r)=(6,2)$ in C and $(7,2)$ in S
have $r-s<-1$. They have $r-s+1=-a_D$; all other singular classes have $r-s\ge0$.

The common missed-jump bond $0\leftrightarrow1$ makes the no-detection evolution positivity improving:
$|0\rangle,|1\rangle$ are jointly cyclic for $H_h$, and one or two missed jumps with
analytic coherent intervals connect every input and output. Thus every row is positive for $t>0$.
The simple Perron exponent and scalar coefficients are even analytic functions of $h$ and hence
analytic in $\theta$, although the generator itself need not be. The uniform spectral gap bounds the
absolute scores by $C_\varepsilon(1+t)$ on $t\ge\varepsilon>0$. Writing $W$ for a single waiting duration, we also have
\begin{equation}
	\gamma I\le D_{\mathrm{det}}\le2\gamma I,\qquad
	\sup_{u,b}\mathbb E_u[\e^{zW}\mid b]\le\frac\gamma{\gamma-z},\quad 0<z<\gamma.
	\label{clock:eq:mgf}
\end{equation}
The distribution $\pi$ samples reset
sites immediately after detections. It is independent of $u$ and has $\pi_4=1/7$, unlike the physical
stationary density matrix $I/8$. To verify invariance, set $d_c=\sum_{a:a\to c}\eta_{ac}$, so
$\sum_cd_c=14$ and $\pi_c=d_c/14$. The sum $\mathcal J$ of registered maps obeys
$\mathcal K_h(I)=-\mathcal J(I)=-\gamma\sum_cd_cP_c$. Hence
$\int_0^\infty\e^{t\mathcal K_h}(\sum_b\pi_bP_b)dt=I/(14\gamma)$. Applying $\mathcal J$ and selecting a
destination proves the claim. For any clock width, the joint invariant law of reset site and within-bin time
offset is $\pi\otimes\mathrm{Uniform}$: an independent uniform offset remains uniform after translation by a
waiting time. Finally,
\begin{equation}
	q_\theta(0\to1,t\mid4)-q_0(0\to1,t\mid4)
	=\frac\gamma2\theta t^2[1+O(t)],\qquad
	q_0(0\to1,t\mid4)\asymp t^{s_D+2}.
	\label{clock:eq:signal}
\end{equation}

\subsection{Regeneration uniform in the clock width}
Fix $0<t_0<t_1$, independently of $\theta,\Delta$. By compactness and
positivity, choose $c>0$ such that
\begin{equation}
	2c\le\inf_{u,b,\,t\in[t_0,t_1]}q_u(3\to4,t\mid b).
	\label{clock:eq:minor}
\end{equation}
The proof tracks the last reset site $b$ and the event's unobserved time offset $v$ within its bin. The state is
$(b,v)\in\{0,\ldots,7\}\times[0,\Delta)$; this time offset is also called the bin phase, and is unrelated to a
quantum phase. A transition with bin increment $k$ and label $\mu$ has wait $w=k\Delta+v'-v>0$ and density
$q_u(\mu,w\mid b)\,dv'$. Define
\begin{equation}
	I_\Delta=\{k:[(k-1)\Delta,(k+1)\Delta]\subset[t_0,t_1]\}.
\end{equation}
On each cell with $\mu=(3\to4)$ and $k\in I_\Delta$, split off a
	constant part $c\,dv'$ and mark that part with an auxiliary flag. Erasing the flag restores the original
law. The remaining kernel $r_u$ is nonnegative and has constant row mass $1-p_\Delta$, where
\begin{equation}
	p_\Delta=c\Delta|I_\Delta|,\qquad
	0<p_*\le p_\Delta\le p^*<1.
	\label{clock:eq:flagprob}
\end{equation}
Choose $\Delta_0$ small enough that these bounds hold for every width at most
$2\Delta_0$, and shrink $\theta_0$ so $\theta_0^{1/s_D}\le\Delta_0$. This also covers the
coarsened receiver width in Sec.~\ref{sec:count-receiver}.

A flagged transition chooses $k$ uniformly on $I_\Delta$, leaves the particle at site $4$,
and has an independent uniform outgoing time offset. All these laws are independent of $u$. With
$K_{\min}=\min I_\Delta$,
\begin{equation}
	\Delta K_{\min}\ge t_0,\qquad \Delta\max I_\Delta\le t_1.
	\label{clock:eq:minadvance}
\end{equation}
This splits the existing transition kernel mathematically; it introduces no physical operation.

In this auxiliary construction, a block contains $n$ residual transitions and ends at the first flag. Assign
the ending flag's outgoing time offset to the next block. With this assignment, blocks retaining their offsets
are independent. The initial block starts at $(4,0)$; later blocks start at site 4 with a uniform offset and
have identical laws. Discarding the offsets retains all directed labels, bin increments, and the ending flag
increment. Let
\begin{equation}
	L=K_{\mathrm{flag}}+\sum_{i=1}^n k_i,\qquad \ell_\Delta=\Delta L\ge t_0.
	\label{clock:eq:blocklength}
\end{equation}
Here $K_{\mathrm{flag}}$ is the bin increment of the ending flag, and $\ell_\Delta$ is a physical bin advance, not the missed fraction.

The number of transitions before a flag is geometric with success probability $p_\Delta$.
Equation~\eqref{clock:eq:mgf} and $\Delta k\le w+\Delta$ give uniform exponential moments of both $n$
and $\sum_i\Delta k_i$. Indeed the residual operator weighted by $\e^{\varepsilon+z\Delta k}$ has row
norm below one for sufficiently small fixed $\varepsilon,z>0$.

A more precise bound retains the small row integral of a selected transition. If a nonnegative one-step function
$A$ has residual row integral at most $A_*$, then, for every fixed integer $d\ge0$,
\begin{equation}
	\mathbb E_u\left[(n+1)^d\sum_{i=1}^n A_i\right]\le C_d A_*.
	\label{clock:eq:momentlemma}
\end{equation}
At fixed $n,i$, integrate the other residual transitions using their constant row mass. The resulting
bound is
\begin{equation}
	p_\Delta A_*\sum_{n\ge1}(n+1)^d n(1-p_\Delta)^{n-1},
\end{equation}
which is uniformly bounded by Eq.~\eqref{clock:eq:flagprob}. Thus the
polynomial factors from the contraction preserve the suppression from the rare short wait.

\subsection{A common bound on the score at finite bin width}
The score can be large for a very short wait, but such waits are rare. We retain both effects and use
Eq.~\eqref{clock:eq:momentlemma} to sum over arbitrarily long blocks. Assume $\theta\le\Delta^{s_D}$ and
$0\le u\le\theta$. A pattern $Y$ retains the block's labels, bin increments, and ending flag increment, with all
time offsets integrated out. We bound its score $\partial_u\log\varpi_u(Y)$, where $\varpi_u$ is the pattern
probability, by one function valid for every such $u$.

For fixed $n$, contract the integrated offsets by $v_j'=(1-\varepsilon)v_j+\varepsilon r_j$, with
$\varepsilon=1/(n+2)$. Along a run of zero bin increments, increase the reference offsets by $\Delta/(n+2)$ at
each step. Restart after a positive increment. The initial reference is zero for the first block and
$\Delta/(n+2)$ for a regular block. On an initial zero-increment run, $r_j=j\Delta/(n+2)$ gives
$w_j'=(1-\varepsilon)w_j+\Delta/(n+2)^2$. In general,
\begin{equation}
	w_i'\ge(1-\varepsilon)w_i,\qquad
	w_i'\ge\frac\Delta{(n+2)^2},\qquad
	|w_i'-w_i|\le2\varepsilon\Delta.
	\label{clock:eq:contract}
\end{equation}
Labels, reset sites, and bin increments are preserved. Each integrated offset contributes a Jacobian factor
$1-\varepsilon$. Polynomial degrees are at most seven, and $n\varepsilon$ is bounded.
Equation~\eqref{clock:eq:rowbounds} controls the remaining logarithmic time derivatives. Together these
estimates bound the product-density and Jacobian loss uniformly in $n,\Delta$. In the subtracted cells, where the flag was split off, $r_u\ge q_u/2$ on $[t_0,t_1]$, so the
logarithmic time and parameter derivatives remain uniformly bounded.

To bound the singular factor, retain
the ratio of its polynomial denominators. Its transformed-to-original ratio is at least a
fixed constant times $(1-\varepsilon)^r[(w_i')^s+u]/[w_i^s+u]$. Change variables, and
enlarge the contracted integration image to the original domain. This gives
\begin{equation}
	\mathbb E_u[(w_i^s+u)^{-1}\mid Y]
	\le C(n+2)^{2s}\Delta^{-s}.
	\label{clock:eq:posterior}
\end{equation}
For $k_i\ge2$, the deterministic bound $w_i\ge(k_i-1)\Delta$ is tighter. Thus the
coarse-pattern score has a common envelope $R_\Delta(Y)$: sum $C(1+\Delta k_i)$ over all
residual transitions, and for each singular row add
\begin{equation}
	\begin{cases}
		C(n+2)^{2s}\Delta^{-s},&k_i\le1,\\
		C[(k_i-1)\Delta]^{-s},&k_i\ge2.
	\end{cases}
	\label{clock:eq:envelope}
\end{equation}
The factor $\Delta k_i$ measures elapsed physical time; $k_i$ counts bins.

For either initial time-offset distribution, the envelope satisfies
\begin{equation}
	\sup_{0\le u\le\theta}\mathbb E_u R_\Delta^2\le C\Delta^{-a_D}.
	\label{clock:eq:envelopemoment}
\end{equation}
Square the sum by Cauchy--Schwarz and absorb its polynomial count factors with
Eq.~\eqref{clock:eq:momentlemma}. For a singular short cell $k\le1$, uniformly in the
incoming state and time offset,
\begin{equation}
	\Pr_u(k\le1,\text{ that row})
	\le C\int_0^{2\Delta}(t^m+u t^r)\,dt\le C\Delta^{m+1},
	\label{clock:eq:shortmass}
\end{equation}
because $u\le\Delta^{s_D}\le\Delta^s$ when $\Delta<1$. Its squared-envelope contribution
is $C\Delta^{m+1-2s}=C\Delta^{r-s+1}$. For $k\ge2$, $w$ and $(k-1)\Delta$ differ by at
most a factor of three, so the contribution is bounded by
\begin{equation}
	C\int_\Delta^\infty q_u(t)t^{-2s}\,dt
	\le C\left[1+\int_\Delta^{\varepsilon_0}t^{r-s}\,dt\right].
	\label{clock:eq:longmass}
\end{equation}
Here $\varepsilon_0>0$ is a fixed short-time cutoff, independent of $\theta$ and $\Delta$, and $u\le t^s$ in this
part; uniform moment bounds control the remaining
integral. The exceptional rows give $\Delta^{-1}$ for C and $\Delta^{-2}$ for S. Every
other singular class is bounded, as are the regular terms. This proves
Eq.~\eqref{clock:eq:envelopemoment}.

Write $\varpi_u(Y)$ for the coarse-block mass. Integrating the score bound gives
$|\log(\varpi_\theta/\varpi_0)|\le\theta R_\Delta$. Integrating the derivative of the mass gives
$|\varpi_\theta-\varpi_0|\le\int_0^\theta\varpi_uR_\Delta\,du$. Substitution into the Jeffreys identity,
	$D_{\mathrm{KL}}(p\Vert q)+D_{\mathrm{KL}}(q\Vert p)=\int(p-q)\log(p/q)$ yields
\begin{equation}
	D_{\mathrm{KL}}(\varpi_\theta\Vert\varpi_0)+D_{\mathrm{KL}}(\varpi_0\Vert\varpi_\theta)
	\le\theta\int_0^\theta\mathbb E_u R_\Delta^2\,du
	\le C\theta^2\Delta^{-a_D}.
	\label{clock:eq:coarsekl}
\end{equation}
The laws have common support. Nonnegativity allows Tonelli's theorem to interchange the
integrals and the sum over the countably many patterns. This bounds the Kullback--Leibler divergence in both
directions without inferring a finite-parameter bound from Fisher information alone.

\subsection{Information within a block and at its boundaries}
Define the nonnegative functions
\begin{equation}
	\Psi(x,y)=x\log(x/y)-x+y,\qquad
	\Phi(x,y)=x^{3/2}y^{-1/2}-\tfrac32x+\tfrac12y.
\end{equation}
They satisfy $\Psi\le2\Phi$ and $\Phi(x,y)\le(x-y)^2/(2y)$ for all $x,y>0$. From the
waiting-density factors for each reset site, for an exceptional row with $r=2$,
\begin{align}
	\Phi(q_\theta,q_0)&\le C\min\{\theta^2t^{2-s_D},\theta^{3/2}t^{2-s_D/2}\},\nonumber\\
	\Phi(q_0,q_\theta)&\le C\frac{\theta^2t^2}{t^{s_D}+\theta}.
	\label{clock:eq:localphi}
\end{align}
Split at $t=\theta^{1/s_D}$. The integrals are $O(\theta^{\nu_D})$; the powers of $t$ in the inner integral are
$t^0$ for C and $t^{-1/2}$
for S, and are integrable. The other singular rows have $r-s\ge0$ and contribute
$O(\theta^2)$, as do regular rows and the positive-time tails.

Since $\Delta k\le t+\Delta$, and subtracted cells are uniformly positive and regular on
a fixed time interval, the residual kernel satisfies
\begin{align}
	\sup_{b,v}\sum_{k,\mu}\int(1+\Delta k)|r_\theta-r_0|\,dv'&\le C\theta,\nonumber\\
	\sup_{b,v}\sum_{k,\mu}\int(1+\Delta k)
	[\Phi(r_\theta,r_0)+\Phi(r_0,r_\theta)]\,dv'&\le C\theta^{\nu_D}.
	\label{clock:eq:weightedrow}
\end{align}
These constants are uniform in $\Delta$. In particular, summing over infinitely many
bin increments introduces no factor proportional to the inverse bin width.

To control an experiment ending inside a bin, we must bound the extra information in the unobserved offsets of one
block. A rich block retains its starting offset and all outgoing offsets of residual transitions, but excludes that
of its ending flag. Its density is $f_u=p_\Delta\beta_\Delta(K_{\mathrm{flag}})\prod_i r_u$, where
$\beta_\Delta$ is uniform on $I_\Delta$. Let $R$ be the residual transition operator and $W$ the same operator
weighted by physical bin advance $\Delta k$. They act on bounded Borel functions of reset site and offset. Set
$H=(I-R)^{-1}$ and $\kappa_\Delta=\mathbb E[\Delta K_{\mathrm{flag}}]$. These are auxiliary kernel operators; $H$ is
not the Hamiltonian $H_h$. The exact $\ell_\Delta$-weighted block mass is, with $\nu$ the joint law of the initial reset site and time offset,
\begin{equation}
	\mathcal F_\nu(R,W)=p_\Delta\nu[\kappa_\Delta H+HWH]\mathbf1.
	\label{clock:eq:functional}
\end{equation}
Here $\nu=\delta_{(4,0)}$ initially or $\delta_4\otimes\mathrm{Uniform}$ later. Ordinary
supremum norms permit the initial point evaluation. Both $\kappa_\Delta$ and the
reciprocal $p_\Delta^{-1}$ are uniformly bounded.

For the ordinary and weighted operator pairs, let
$\mathbf D=\mathbf R_\theta-\mathbf R_0$. The two tilted kernels $r_\theta^{3/2}r_0^{-1/2}$ and
$r_0^{3/2}r_\theta^{-1/2}$ give
\begin{equation}
	\mathbf B_+=\mathbf R_0+\tfrac32\mathbf D+\mathbf E_+,
	\qquad \mathbf B_-=\mathbf R_0-\tfrac12\mathbf D+\mathbf E_-,
	\qquad \|\mathbf D\|=O(\theta),\quad \|\mathbf E_\pm\|=O(\theta^{\nu_D}).
	\label{clock:eq:tilt}
\end{equation}
Their ordinary row norms are $1-p_\Delta+O(\theta^{\nu_D})<1$ uniformly, because $\mathbf D$ has zero ordinary row
mass. The first and second derivatives of Eq.~\eqref{clock:eq:functional} are uniformly bounded. Homogeneity of
degree one in $(x,y)$ gives
\begin{align}
	\int\ell_\Delta\Phi(f_\theta,f_0)
	&=\mathcal F_\nu(\mathbf B_+)-\tfrac32\mathcal F_\nu(\mathbf R_\theta)+\tfrac12\mathcal F_\nu(\mathbf R_0),\nonumber\\
	\int\ell_\Delta\Phi(f_0,f_\theta)
	&=\mathcal F_\nu(\mathbf B_-)-\tfrac32\mathcal F_\nu(\mathbf R_0)+\tfrac12\mathcal F_\nu(\mathbf R_\theta).
\end{align}
Taylor expansion cancels the constant and linear $\mathbf D$ terms in both combinations, including the perturbation
of the physically weighted operator. The remaining terms give
\begin{equation}
	\int\ell_\Delta[\Psi(f_\theta,f_0)+\Psi(f_0,f_\theta)]\le C\theta^{\nu_D}.
	\label{clock:eq:rich}
\end{equation}
The same argument with the unweighted functional $p_\Delta\nu H\mathbf1$ bounds the rich-block Jeffreys divergence
by $C\theta^{\nu_D}$.

Let $g_+(Y),g_-(Y)$ denote the forward and reverse KL divergences of the conditional law of the retained time
offsets given a coarse block. Conditioning on the coarse block gives the exact decomposition
\begin{equation}
	\int\ell_\Delta\Psi(f_\theta,f_0)
	=\sum_Y\ell_\Delta\Psi(\varpi_\theta(Y),\varpi_0(Y))
	+\mathbb E_\theta[\ell_\Delta g_+(Y)],
\end{equation}
and the reversed identity. The coarse terms are nonnegative, so
\begin{equation}
	\mathbb E_\theta[\ell_\Delta g_+]+\mathbb E_0[\ell_\Delta g_-]
	\le C\theta^{\nu_D}.
	\label{clock:eq:conditional}
\end{equation}
This bound is uniform and weighted by the elapsed time of the one block whose time offsets may be needed at an
arbitrary endpoint.

\subsection{Lower bound for an experiment ending at any fixed time}
Choose $N=\lfloor T/t_0\rfloor+1$. By Eq.~\eqref{clock:eq:minadvance}, the first $N$ blocks
already extend beyond $T$ in bin position, and $N\le C(T+1)$. At a full-bin horizon, their
coarse prefix determines the entire observed record. In the coarse regime,
the additivity of the KL divergence over independent
blocks and Eq.~\eqref{clock:eq:coarsekl} give $C(T+1)\theta^2\Delta^{-a_D}$.

For arbitrary $T=(M+\rho)\Delta$, $0\le\rho<1$, put
\begin{equation}
	S_j=\sum_{i=1}^jL_i,\qquad J=\min\{j:S_j>M\}\le N.
\end{equation}
The coarse prefix determines $J$. Reveal only block $J$'s retained time offsets. Earlier residual events lie at
least $K_{\min}$ bins before their ending flags. If the previous flag falls in bin $M$, its outgoing offset
belongs to block $J$. These offsets therefore locate all registered events in bin $M$; the ending flag of $J$
and all later blocks lie beyond it. At $M=0$, the supplied reset is not a registered event. The enriched record
thus reconstructs the true partial-bin observation, including final silence.

Independence and the assignment of offsets imply that, conditional on the entire coarse prefix, the selected rich
law is precisely block $J$'s conditional law given $Y_J$. The endpoint-selected block need not have an unselected
block's law. For a regular block, we control this selection by counting possible start bins in
$[M-L(y)+1,M]$. Integrating $g_\pm(y)$ against their renewal occupation
	measure, the expected number of block starts falling in each bin interval, gives the expected conditional KL.
Starts are separated by at least $K_{\min}$ bins, so this interval contains at most
\begin{equation}
	1+\frac{L(y)}{K_{\min}}\le\frac{2L(y)}{K_{\min}}
	\le\frac{2}{t_0}\ell_\Delta(y)
	\label{clock:eq:occupation}
\end{equation}
starts, since $L(y)\ge K_{\min}$. The contribution from a selected regular block is
therefore at most $(2/t_0)\mathbb E[\ell_\Delta g_\pm]$. The initial block contributes at
most $t_0^{-1}\mathbb E[\ell_\Delta g_\pm]$. The KL chain rule and
Eq.~\eqref{clock:eq:conditional} prove, in the coarse regime,
\begin{equation}
	D_{\mathrm{KL}}(P_\theta\Vert P_0)+D_{\mathrm{KL}}(P_0\Vert P_\theta)
	\le C(T+1)\theta^2\Delta^{-a_D}+C\theta^{\nu_D}.
	\label{clock:eq:coarsehorizon}
\end{equation}
No other block's time offsets are revealed in this coarse-regime argument.

In the fine regime $\Delta\le\theta^{1/s_D}$, reveal all retained time offsets of the same deterministic $N$-block
prefix. Its last ending flag is beyond $T$, so that flag's omitted outgoing time offset is irrelevant. The
unweighted rich-block bound gives $C(T+1)\theta^{\nu_D}$ even for arbitrary endpoints. Combining the two regimes
proves Eq.~\eqref{clock:eq:main}, because
\begin{equation}
	\nu_D=2-\frac{a_D}{s_D},\qquad
	\min\{\theta^2\Delta^{-a_D},\theta^{\nu_D}\}
	\asymp\frac{\theta^2}{(\Delta+\theta^{1/s_D})^{a_D}}.
\end{equation}
For full-bin horizons, the coarse prefix alone suffices in the coarse regime, so the additive term is absent. This
argument never replaces a partial-bin record by a rounded-up coarse record through data processing.

If a test has equal-prior error at most $0<e<1/2$, its two observation laws have total variation at least $1-2e$.
Pinsker's inequality and Eq.~\eqref{clock:eq:main}, with $\theta^{\nu_D}\to0$, therefore imply $T\ge c_e/R_D$ for
sufficiently small $\theta$. The bound is uniform in the joint clock limit.

The matching upper time bound in Theorem~\ref{clock:thm:frontier} follows from the counting rule in
Sec.~\ref{sec:count-receiver}, which uses a deterministic full-bin endpoint. It is therefore also a valid test on
the richer record bounded here. Together, the lower bound and the counting rule give the stated time scalings.
The splitting used in the proof introduces no physical reset.
\section{Detection time with missed jumps and finite time resolution}
\label{sec:oct18-joint}
\label{sec:count-receiver}
\subsection{Experiment, domain, and results}

On sites $j=0,\ldots,7$ modulo eight, let $P_j=|j\rangle\langle j|$ and
\begin{align}
	H_h&=\sum_{j=0}^7(|j+1\rangle\langle j|+\mathrm{H.c.})
	+h(|0\rangle\langle4|+|4\rangle\langle0|),\\
	L_{a\to c}&=\sqrt\gamma|c\rangle\langle a|,\qquad c=a\pm1.
\end{align}
Frequencies are in units of $J$ and times in units of $J^{-1}$. The rate
$\gamma>0$ is fixed and known. Arrangement C has a missed fraction of $1/2$ in both
directions of edges $0$--$1$ and $1$--$2$; S has it on edges $7$--$0$ and $0$--$1$. The other
twelve directed channels have a common missed fraction $\ell$. The detector does not affect the
unconditional generator $\mathcal L_h$. The particle is prepared once in $P_4$.

The hypotheses are $h=0$ and one specified $h>0$. In each absolute hardware bin of width $\Delta$,
retain only the numbers of events in
\begin{equation}
	F=(3\to4),\qquad A=(0\to1),\qquad O=\text{other registered events}.
\end{equation}
No sub-bin times or within-bin order are kept. At a deterministic horizon $T$, the final partial bin,
if any, is censored at $T$. Put
\begin{equation}
	\theta=h^2,\quad p=\begin{cases}2&\mathrm C,\\3&\mathrm S,\end{cases}
	\quad s=p+2,\quad \rho=\theta^{1/s},\quad L=\ell+\Delta+\rho.
	\label{oct18-joint:eq:scales}
\end{equation}
Let $T^\star_{e,D,\mathrm{count}}(h,\ell,\Delta)$ be the minimum deterministic horizon for equal-prior error at
most $e$, optimized over all decisions using this per-bin count-vector record. It is not restricted to a total
pattern count. The criterion requiring each error to be at most $e$, used for $T_e^D$ in the Letter, yields the same time order by the argument following Eq.~\eqref{eq:risk}. There is no
feedback, external reset, dead time, dark count, or label error.

\begin{theorem}[Minimum time with missed jumps and finite bin width]\label{oct18-joint:thm:joint}
	Fix $\gamma>0$ and equal-prior error $e\in(0,1/2)$. There are positive
	$h_0,\ell_0,\Delta_0$ and constants independent of $h,\ell,\Delta$ such that, for
	$0<h\le h_0$, $0\le\ell\le\ell_0<1/2$, and $0<\Delta\le\Delta_0$,
	\begin{equation}
		T^\star_{e,D,\mathrm{count}}(h,\ell,\Delta)
		\asymp \theta^{-2}(\ell+\Delta+\rho)^{p-1}.
		\label{oct18-joint:eq:joint}
	\end{equation}
	The same order holds for any record that contains these counts and can be obtained from
	the full ordered record of directed labels and absolute bin indices.
\end{theorem}
The parameter bounds are chosen so that $\ell_0+\Delta_0+h_0^{2/s}$ lies in a single short-time window. Arbitrarily coarse clocks are not covered. A fixed positive
$\ell$ or clock width inside this rectangle gives $h^{-4}$. The constants
are not uniform as $\gamma\to0$ or $\gamma\to\infty$.

\subsection{Lower bounds before combining detector labels}

The positive-$\ell$ exact-time theorem in Sec.~\ref{sec:optimal},
Eq.~\eqref{eq:optimal-time}, and the zero-$\ell$ absolute-clock theorem in
Sec.~\ref{sec:absolute-clock} supply, respectively,
\begin{equation}
	T\ge c_e\theta^{-2}(\ell+\rho)^{p-1},\qquad
	T\ge c'_e\theta^{-2}(\Delta+\rho)^{p-1}.
	\label{oct18-joint:eq:boundaries}
\end{equation}
Their statements are uniform in a small parameter range and include
arbitrary deterministic endpoints.

Exact timestamps contain more information than the positive-$\ell$ count record. For the clock bound, start from
the zero-$\ell$ directed-label binned record and delete every protected-channel report independently with
probability $\ell$. Keep the four high-loss channels unchanged. A deleted report does not undo the physical jump.
Summing over deleted outcomes gives exactly the positive-$\ell$ counting experiment, including its conditional state updates, since no
feedback changes the dynamics. This rule is independent of the hypothesis at every horizon, including partial
bins. Only after deletion are surviving labels pooled into $F,A,O$.

It is generally impossible to implement this deletion using the zero-$\ell$
$O$ count alone: $O$ mixes protected and high-loss channels. The lower-bound argument uses the
richer directed-label record covered by the established theorem. Data processing therefore gives
both inequalities in Eq.~\eqref{oct18-joint:eq:boundaries} for the joint count experiment. Since
\begin{equation}
	\max(\ell+\rho,\Delta+\rho)\le L
	\le2\max(\ell+\rho,\Delta+\rho),
\end{equation}
their maximum gives the lower bound in Eq.~\eqref{oct18-joint:eq:joint}. The combination follows from two distinct
loss-of-information arguments, not from replacing missed-jump recycling by time binning.

\subsection{Two counting intervals at nonzero missed fraction}

Choose
\begin{equation}
	k=\max\{1,\lceil(\ell+\rho)/\Delta\rceil\},\qquad W=k\Delta.
	\label{oct18-joint:eq:W}
\end{equation}
Then $\max\{\Delta,\ell+\rho\}\le W\le L$, $\ell\le W$, and $\theta\le W^{p+2}$. Sum the
count vectors in $k$ consecutive hardware bins to obtain the counts in an interval of
width $W$. A disjoint pair is successful only when its vectors are $(1,0,0)$ and
$(0,1,0)$ in this order.

The no-registration generator accounts for every missed-jump history:
\begin{equation}
	\mathcal K_{h,\ell}X=-i[H_h,X]-2\gamma X
	+\gamma\sum_{a\to c}\delta_{ac}X_{aa}P_c.
\end{equation}
The actual registered maps are
\begin{equation}
	\mathcal J_FX=\gamma(1-\ell)X_{33}P_4,\qquad
	\mathcal J_AX=\frac\gamma2 X_{00}P_1.
\end{equation}
The map for a successful pair is exactly
\begin{equation}
	\mathcal E_{h,\ell}^{W}=\int_0^Wdu\int_W^{2W}dv\,
	\e^{(2W-v)\mathcal K}\mathcal J_A
	\e^{(v-u)\mathcal K}\mathcal J_F\e^{u\mathcal K}.
	\label{oct18-joint:eq:E}
\end{equation}
Here $u$ is the time of the $F$ jump within the first interval and $v$ the time of the $A$ jump within
	the second. The expression incorporates evolution with no detection before, between,
and after the two registered events. The $F$ jump leaves the particle in $P_4$, but missed jumps can
move it again before the pair ends. The state at the pair boundary is
therefore generally mixed. All unselected outcomes sum to the completely positive map
$\e^{2W\mathcal L_h}-\mathcal E$.

For positive input $R$, set
\begin{align}
	f_h(u;R)&=\gamma(1-\ell)\langle3|\e^{u\mathcal K}R|3\rangle,\\
	q_h(t)&=(\gamma/2)\langle0|\e^{t\mathcal K}P_4|0\rangle,\\
	S_h(r)&=\Tr\e^{r\mathcal K}P_1.
\end{align}
Here $f_h$ gives the initial $F$ flux, $q_h$ the density for the next detection to be $A$, and $S_h$ the final
no-detection probability. The rank-one jumps split the evolution into three
	factors, giving $\Tr\mathcal E R=\int\!\int f_hq_hS_h$. The target-density bounds are
\begin{equation}
	q_0(t)\asymp t^4(\ell+t)^p,\qquad
	q_h(t)-q_0(t)=\frac\gamma2\theta t^2[1+O(t)].
	\label{oct18-joint:eq:target}
\end{equation}
The physical stationary state is $\sigma=I/8$. The sign symmetry of the counting probabilities makes
$f_h(u;\sigma)$ and $S_h(r)$ even analytic functions of $h$. Uniformly at short times,
\begin{equation}
	f_h(u;\sigma)=\gamma(1-\ell)/8+O(u),\quad S_h(r)=1+O(r),
	\quad f_h-f_0=O(\theta),\quad S_h-S_0=O(\theta).
\end{equation}
Decompose the difference $f_hq_hS_h-f_0q_0S_0$ as
\begin{equation}
	f_h(q_h-q_0)S_h+(f_h-f_0)q_0S_h+f_0q_0(S_h-S_0).
\end{equation}
Using $\int_0^Wdu\int_W^{2W}dv\,(v-u)^2=7W^4/6$ gives
\begin{equation}
	m_h-m_0=\frac{7\gamma^2(1-\ell)}{96}\theta W^4[1+O(W)],
	\qquad m_h=\Tr(\mathcal E_{h,\ell}^W\sigma).
	\label{oct18-joint:eq:separation}
\end{equation}
The terms containing the initial-flux and final-survival differences are
$O(\theta W^{p+6})$ for $\ell\le W$. Thus missed jumps at nonzero $\ell$ do not
cancel the leading signal. All the following local estimates remain valid for $\ell\le2W$, with
modified constants.

Complete positivity, trace nonincrease, and Eq.~\eqref{oct18-joint:eq:target} imply
\begin{align}
	m_0&\asymp W^{p+6},\\
	0\le\Tr(\mathcal E_{h,\ell}^WR)
	&\le C\{W^6(\ell+W)^p+\theta W^4\}\Tr R
	\le A_W\Tr R,\qquad A_W=C_0W^{p+6}.\label{oct18-joint:eq:rarity}
\end{align}
Equivalently, $0\le\mathcal E^*(I)\le A_WI$, uniformly for arbitrary incoming
states. This bound on the pattern probability accounts for all missed-jump
recycling at nonzero $\ell$.

\subsection{Correlations, the initial state, and observation time}

Write $\mathcal L_h=-i[H_h,\cdot]+\mathcal D$. The physical dissipator acts as
\begin{equation}
	\mathcal D X=\gamma\sum_jX_{jj}(P_{j+1}+P_{j-1})-2\gamma X.
\end{equation}
On diagonal matrices, the Fourier modes have eigenvalues $-2\gamma[1-\cos(2\pi k/8)]$,
with $k=0,\ldots,7$. On off-diagonal matrices, the eigenvalue is $-2\gamma$. The
dissipator is self-adjoint in the Hilbert--Schmidt inner product and its smallest
decay rate on traceless matrices is $\lambda=2\gamma[1-\cos(\pi/4)]$. The Hamiltonian commutator contributes nothing to the
	derivative of the squared norm. Differentiating the squared norm therefore
gives the following bound for every traceless $Y$,
\begin{equation}
	\norm{\e^{t\mathcal L_h}Y}_2\le\e^{-\lambda t}\norm Y_2.
	\label{oct18-joint:eq:mix}
\end{equation}
This physical generator is independent of the missed fraction.

For any initial density matrix $R$, write $\Phi=\e^{2W\mathcal L_h}$ and $E=\mathcal E_{h,\ell}^W$. Before pair
$i$ the unconditional state is $R_i=\Phi^{i-1}R$, with success probability $p_i=\Tr ER_i$; its stationary value is
$m=\Tr E\sigma$. Set $Y_i=ER_i-p_i\Phi R_i$. This centered matrix has $\Tr Y_i=0$ and $\norm{Y_i}_1\le2p_i\le2A_W$.
Instrument composition gives
\begin{align}
	\operatorname{Cov}(X_i,X_j)&=\Tr(E\Phi^{j-i-1}Y_i),\\
	|\operatorname{Cov}(X_i,X_j)|&\le2\sqrt8 A_W^2
	\e^{-2\lambda W(j-i-1)}.
\end{align}
The covariance contains two factors of the rare pattern probability, one from each selected pair. Summation and
Eq.~\eqref{oct18-joint:eq:mix} imply
\begin{align}
	\operatorname{Var}Z_n&\le Cn(A_W+A_W^2/W)\le CnW^{p+6},\label{oct18-joint:eq:var}\\
	|\E Z_n-nm|&\le CA_W/W\le CW^{p+5},\qquad Z_n=\sum_{i=1}^nX_i.
	\label{oct18-joint:eq:startup}
\end{align}
These bounds hold for every $R$, including a state conditioned on an earlier
calibration record.

Choose $n=\lceil M\theta^{-2}W^{p-2}\rceil$ with fixed multiplier $M>0$. The count threshold $n(m_0+m_h)/2$ lies
midway between stationary means separated by at least $cn\theta W^4$. The initial-state bias relative to this
separation is at most $C\theta W^3/M$. Chebyshev's inequality then bounds each error by $C/M$. Observation ends
at $T=2nW=2nk\Delta$, so
\begin{equation}
	T\le C\theta^{-2}W^{p-1}\asymp\theta^{-2}L^{p-1}.
\end{equation}
The ceiling adds only $2W$. Empty pairs, rejected pairs, missed returns,
and final silence are all included. This completes Theorem~\ref{oct18-joint:thm:joint}.

\section{Finite error at matched detector settings}
\label{sec:finite-example}
\label{sec:oct18-gap}
The comparison in the Letter fixes $\gamma=1$, $\ell=1/10000$, $\Delta=1/25$, and the
alternative $h=1/1000$. Both arrangements are prepared once in $P_4$, use the same
absolute clock, and record the same three classes. The efficiencies remain fixed under the two
hypotheses. Their stationary collection fraction is $34997/40000=0.874925$. An initial interval of
20, or 500 hardware bins, is followed by $M$ disjoint five-bin blocks. The total time is $T=20+M/5$.

\subsection{One common count statistic}
In this finite-parameter section, $\mathcal K_h$ abbreviates $\mathcal K_{h,\ell}$ at $\ell=10^{-4}$; the
generator in Sec.~\ref{sec:absolute-clock} instead has $\ell=0$. Let $Z_h=\e^{\Delta\mathcal K_h}$ describe a
bin with no registration. The map for exactly one
	$F$ and no other registration is
\begin{equation}
	F_h=\int_0^\Delta \e^{(\Delta-u)\mathcal K_h}\mathcal J_F \e^{u\mathcal K_h}\,du,
	\qquad \mathcal J_FX=(1-\ell)X_{33}P_4.
\end{equation}
Replace $\mathcal J_F$ by $\mathcal J_AX=X_{00}P_1/2$ to define $A_h$. Count the pattern
$(F,0,0,0,A)$, where each zero means a bin with no registered event. Its map and
stationary probability are
\begin{equation}
	B_h^{\rm event}=A_hZ_h^3F_h,\qquad
	\mu_h=\Tr B_h^{\rm event}\pi,\qquad \pi=I/8.
	\label{eq:oct18-fivebin}
\end{equation}
Summing all outcomes in a block gives the map $\Phi_h=\e^{5\Delta\mathcal L_h}$. Both
$B_h^{\rm event}$ and $\Phi_h-B_h^{\rm event}$ are completely positive because they sum
disjoint physical records. The latter covers every rejected pattern,
including patterns with registered events. All five bins are counted. Let
$N_M$ be the total number of accepted blocks. Its probability law retains correlations between
blocks.

\subsection{Probability bounds at the specified link strength}
Represent the Hermitian density matrix by 64 real coordinates: its eight populations, the
real parts of its 28 upper off-diagonal entries, and their imaginary parts. Write the
$64\times64$ matrix of $\mathcal K_h$ in these coordinates as $K_h$. The augmented matrix
\begin{equation}
	G_F=\begin{pmatrix}K_h&0\\J_F&K_h\end{pmatrix}
\end{equation}
propagates the upper sector without a registered event. One registered $F$ transfers it to the lower sector,
which then evolves without further registration. Extracting the second sector of $\e^{\Delta G_F}(x,0)$
therefore gives exactly one $F$ event, $F_hx$. No other registered jump enters either diagonal block. Replacing
$J_F$ gives $A_h$, while $Z_h$ uses $K_h$ alone. All physical missed jumps remain in $K_h$.

Under both hypotheses, multiplying these scaled matrices by 250000 gives integer entries. The induced
infinity norm of each scaled matrix is at most $q=3/10$ in both arrangements. For any such matrix $Q$, use
\begin{equation}
	P_{64}(Q)=\sum_{j=0}^{64}\frac{Q^j}{j!},\quad
	s(q)=\sum_{j=0}^{64}\frac{q^j}{j!},\quad
	t(q)=\frac{q^{65}}{65!(1-q/66)}.
	\label{eq:finite-exp-tail}
\end{equation}
Then $\|\e^Q-P_{64}(Q)\|_\infty\le t(q)$. If the current polynomial vector has norm bound
$B$ and error bound $E$, the next bin has bounds
\begin{equation}
	E_{\rm new}\le[s(q)+t(q)]E+t(q)B,\qquad
	B_{\rm new}\le s(q)B.
\end{equation}
Extracting the required sector has norm one. The input norm is $1/8$, and the final trace costs a factor at most
eight. These cancel, so scalar propagation starts at $B=1,E=0$. Five applications bound the final probability
error by $4\times10^{-124}$. Integer Taylor recurrence and rational error bounds give the outward-rounded
intervals in Table~\ref{tab:finite-probabilities}; each entry is multiplied by $10^{-12}$:

\begin{table}[!htbp]
	\centering
	\caption{Enclosures of dimensionless stationary five-bin pattern probabilities at $\gamma=1$, $\ell=10^{-4}$,
		$\Delta=0.04$, and $h=0$ or $0.001$.}
	\label{tab:finite-probabilities}
	\begin{tabular}{ccc}
		\toprule
		Arrangement & $10^{12}\mu_0$ & $10^{12}\mu_{0.001}$\\
		\midrule
		C & $[3.693713139866,\,3.693713139867]$ & $[5.398386730880,\,5.398386730881]$\\
		S & $[1.133908952850,\,1.133908952851]$ & $[2.824465473261,\,2.824465473262]$\\
		\bottomrule
	\end{tabular}
\end{table}

Thus S suppresses the null pattern probability by a factor of about $3.26$, while
$(\mu_h-\mu_0)_{\rm S}/(\mu_h-\mu_0)_{\rm C}\simeq0.9917$. The Python program
\texttt{certify\_matched.py} uses only the standard library, propagates the density matrix
at both signal strengths, and saves the rational bounds. No expansion in $h$ or trajectory sampling enters
these bounds. The program and saved data are available from the corresponding author upon reasonable request.

\subsection{Correlated-count risk and the common horizon}
For maps $\mathcal S$ and $\mathcal T$, write $\mathcal S\gecp\mathcal T$ when
$\mathcal S-\mathcal T$ is completely positive. For either hypothesis, $\mu_h<10^{-10}$.
Fifty blocks span a time of ten. Applying the complement criterion of
Sec.~\ref{sec:oct18-poisson} gives
\begin{equation}
	(\Phi_h-B_h^{\rm event})^{50}\gecp
	\left[1-\frac{64}{148}-\frac{512\cdot50}{10^{10}}\right]D
	>\tfrac12D,\qquad D(X)=\Tr(X)I/8.
\end{equation}
The rare-count lemma, with $m=1/8$, $K=50$, and $c=1/2$, therefore bounds the
stationary count law's total-variation distance from $\operatorname{Pois}(M\mu_h)$ by
$2700\mu_h$, uniformly in $M$. The initial interval adds $b=3/40000$ for the entire
subsequent record, as shown below.

For each hypothesis $i\in\{0,1\}$, enclose the exact probability in
an interval with rational center $\bar\mu_i$ and radius $r_i$. The midpoint and half-width of each
printed interval suffice for the stated risk bounds. A coupling of the two
	Poisson laws bounds the effect of changing the mean by $M r_i$. Hence, with $\mu_i^+$ denoting the upper
endpoint,
\begin{equation}
	\TV(\mathcal L_{P_4}(N_M\mid i),Q_i)\le\epsilon_i,
	\qquad Q_i=\operatorname{Pois}(M\bar\mu_i),\qquad
	\epsilon_i=b+2700\mu_i^++Mr_i.
	\label{eq:finite-TV}
\end{equation}
The minimum equal-prior error from this scalar count is
$R_D^{\rm count}=(1-\TV(P_0,P_1))/2$, where $P_i$ denotes the actual
law of the count. The triangle inequality gives
\begin{equation}
	|R_D^{\rm count}-R_D^{\rm Pois}|\le(\epsilon_0+\epsilon_1)/2.
	\label{eq:finite-Bayes}
\end{equation}
Here $R_D^{\rm Pois}=(1-\TV(Q_0,Q_1))/2$. The estimate bounds the optimum count error from both sides. Any fixed
decision's average error changes by at most $(\epsilon_0+\epsilon_1)/2$; its false-alarm and missed-link
probabilities change by at most $\epsilon_0$ and $\epsilon_1$, respectively. Thus the optimal reference Poisson
threshold gives an explicit rule attaining the stated upper bound for the actual correlated count.

Put $\lambda_i=M\bar\mu_i$. Since $\lambda_1>\lambda_0$, the Poisson likelihood ratio
increases with count. The rule that declares a
link when $n\ge r$ is optimal when
\begin{equation}
	(\lambda_1/\lambda_0)^{r-1}\le \e^{\lambda_1-\lambda_0}
	\le(\lambda_1/\lambda_0)^r.
\end{equation}
The two Poisson errors are the false-alarm probability $1-\e^{-\lambda_0}\sum_{j=0}^{r-1}
\lambda_0^j/j!$ and the missed-link probability $\e^{-\lambda_1}\sum_{j=0}^{r-1}\lambda_1^j/j!$.
The positive exponential is bounded by its
degree-256 Taylor sum and the geometric remainder as in Eq.~\eqref{eq:finite-exp-tail};
reciprocals then bound the negative exponential. All bounds and the likelihood crossing are
checked with rational arithmetic.

At $M=5.5\times10^{12}$, the common total duration is $1\,100\,000\,000\,020$. Table~\ref{tab:finite-risks} gives
the outward-rounded risk and error intervals.

\begin{table}[!htbp]
	\centering
	\caption{Bounds for the optimum equal-prior scalar-count risk and for the two conditional errors of threshold
		$r$, at $M=5.5\times10^{12}$. Counts are integers and risks are dimensionless probabilities.}
	\label{tab:finite-risks}
	\begin{tabular}{ccccc}
		\toprule
		Arrangement & $r$ & $R_D^{\rm count}$ & False alarm at $r$ & Missed link at $r$\\
		\midrule
		C & 25 & $[0.172779,0.172930]$ & $[0.174825,0.174976]$ & $[0.170732,0.170883]$\\
		S & 11 & $[0.073866,0.074017]$ & $[0.053098,0.053249]$ & $[0.094634,0.094785]$\\
		\bottomrule
	\end{tabular}
\end{table}

At this horizon, $Mr_i=2.75\times10^{-12}$ for each printed probability interval, and
$2700\mu_i^+<1.46\times10^{-8}$. The interval width is therefore set mainly by the one-time initial-state bound
$b=7.5\times10^{-5}$, rather than by matrix truncation or rounding. The two conditional errors need not be equal
because the threshold minimizes their average. Thus S attains both errors below $0.1$, while C cannot attain both
from this same scalar count. Each threshold uses the known calibration for its arrangement; both arrangements use
the same observation rule, elapsed time, and hardware resources. These bounds do not
compare optimal decisions from the complete finite-time record. Figure~2(b) of the Letter uses the same
fixed parameters and probability enclosures at twenty different horizons. The source is
prepared once, and neither the proof decomposition nor the receiver supplies an external
reset.

\section{Poisson approximation for counts of rare event sequences}
\label{sec:oct18-poisson}
The finite-time bound uses a Poisson approximation for the number of selected five-bin
patterns. The argument retains correlations, the initial preparation, and
the final censoring. Limits for pattern counts have been used earlier for
quantum output inference~\cite{GirottiGodleyGuta2025};
the comparison here applies directly to the original counting record at finite signal strength.

\subsection{Count bound for a full-rank stationary state}
Let $\Phi$ be a finite-dimensional completely positive, trace-preserving (CPTP) map with full-rank stationary
state $\pi$, $\Tr\pi=1$. Let $A$ and $B=\Phi-A$ be completely positive (CP). They describe the selected and
rejected outcomes of each disjoint block. A selected event may be a union of outcomes or an exact finite
sequence; the entire block is counted. Put
\[
\mu=\Tr A(\pi),\qquad D_\pi(X)=\Tr(X)\pi,\qquad
m=\lambda_{\min}(\pi)>0,
\]
and let $N_M=\sum_{i=1}^MY_i$ count the $A$ outcomes from stationary input, with $Y_i$ the indicator that
	block $i$ yields the selected event. A zero indicator
$Y_i=0$ means that the selected event did not occur. It does not mean zero physical jumps
or zero registered counts. In particular, $B$ differs from the physical
no-registration map $Z$.

\begin{lemma}[Uniform rare-count approximation]
	Suppose that for some integer $K\ge1$ and $0<c<1$,
	\begin{equation}
		B^K\gecp cD_\pi.
		\label{eq:oct18-complement-floor}
	\end{equation}
	Then, uniformly for every $M\ge1$,
	\begin{equation}
		\TV(\mathcal L_\pi(N_M),\operatorname{Poisson}(M\mu))
		\le(1-\e^{-M\mu})\frac{(1+1/m)K(2/c-1)}{1-c}\mu.
		\label{eq:oct18-poisson-general}
	\end{equation}
\end{lemma}
When $\mu=0$, full rank and positivity imply $A=0$, and the assertion
is immediate. Assume $\mu>0$ below.

\begin{proof}
	First, positivity gives
	$\|A^*(I)\|\le\Tr A^*(I)\le\mu/m$.
	We use the stationary reversed instrument to bound events before a selected occurrence. The stationary reverse of
	a CP map is
	\begin{equation}
		T^{\mathrm R}(X)=\pi^{1/2}T^*(\pi^{-1/2}X\pi^{-1/2})\pi^{1/2}.
		\label{eq:oct18-reversal}
	\end{equation}
	The reversed aggregate $\Phi^{\mathrm R}$ is trace preserving because its effect is
	$\pi^{-1/2}\Phi(\pi)\pi^{-1/2}=I$. It fixes $\pi$ because $\Phi^*(I)=I$. The reversed event effect
	$(A^{\mathrm R})^*(I)=\pi^{-1/2}A(\pi)\pi^{-1/2}$ obeys the same norm bound $\mu/m$. Reversal leaves the reset map $D_\pi$ unchanged, is involutive, and reverses composition order. To verify the record
	identity, set $S_\pi(X)=\pi^{1/2}X\pi^{1/2}$, so $S_\pi(I)=\pi$ and $S_\pi^{-1}(\pi)=I$. Intermediate factors
	$S_\pi^{-1}S_\pi$ cancel in the product of reverse maps. Taking the trace and using $\Tr[S_\pi(X)]=\Tr(\pi X)$ gives
	\begin{equation}
		\Tr[T_n\cdots T_1(\pi)]
		=\Tr[T_1^{\mathrm R}\cdots T_n^{\mathrm R}(\pi)].
		\label{eq:oct18-reverse-record}
	\end{equation}
	This is stationary operation reversal~\cite{Crooks2008}, not a
	detailed-balance assumption.
	
	Group the blocks into sets of $K$. Split the all-zero-indicator outcome $B^K$
	into $cD_\pi$ and $B^K-cD_\pi$. Erasing this proof flag gives back the original binary record.
	The flag probability is $c$, independent of the incoming state. Put $q=1-c$
	and $qQ=\Phi^K-cD_\pi$; $Q$ is CPTP and fixes $\pi$. Conditionally on any prescribed flag/no-flag pattern, each
	normalized aggregate preserves $\pi$. The stationary two-sided flag process is therefore independent
	Bernoulli$(c)$, and cycles between flags are independent given the pattern. The same construction applies to the
	reversed instrument.
	
	Choose an index $i$ uniformly from $1,\ldots,M$ and condition on $Y_i=1$. Averaging over the index weights a count
	$n$ in proportion to $n$, giving the size-biased count law $N_M^s$. The central group containing $i$ cannot be
	flagged; outer flags retain their independent Bernoulli law.
	This holds jointly for any finite outer flag pattern: normalized aggregates before the
	event preserve $\pi$, and subsequent flag probabilities are state independent. Couple
	conditioned and unconditioned processes using common outer flags; the unconditioned central
	flag is an independent Bernoulli$(c)$. Let $L$ count the groups between the nearest common flags on the left and
	right. It includes the central group but excludes the two flagged boundary groups. Each side has a geometric
	number of unflagged groups with mean $q/c$, so $\mathbb EL=1+2q/c=2/c-1$. Couple all cycles outside this region
	identically. A central flag in the unconditioned process only splits the region into two cycles.
	
	For position $j$ in a $K$-block group, the unconditioned event map is $A_j=\Phi^{K-j}A\Phi^{j-1}$. Conditioning
	on an unflagged group divides it by $q$. Its stationary mean is therefore $\mu/q$, with forward and reversed
	effect bounds $\mu/(mq)$, since channels contract positive effects in operator norm. Conditioned on the central
	event, every other indicator in a different unflagged group has probability
	at most $\mu/(mq)$. The same bound for earlier events follows from
	Eq.~\eqref{eq:oct18-reverse-record}. Two distinct events in the same
	group have stationary joint probability at most $\mu^2/m$, so their
	conditional probability is at most $\mu/m$, which is stronger.
	The expected count in the affected region for the unconditioned process is at most
	$KL\mu/q$; the conditioned count excluding its forced event is at most
	$KL\mu/(mq)$. Thus a coupling satisfies
	\begin{equation}
		\mathbb E|N_M+1-N_M^s|
		\le(1+1/m)K(\mu/q)(2/c-1).
	\end{equation}
	The size-bias Poisson Stein bound~\cite{ChenFangShao2013} is
	\[
	\TV(\mathcal L(W),\operatorname{Poisson}(\lambda))
	\le(1-\e^{-\lambda})\mathbb E|W+1-W^s|.
	\]
	For completeness, the indicator-test Stein solution satisfies $\|\Delta f\|_\infty\le(1-\e^{-\lambda})/\lambda$.
	The size-bias identity changes the Stein residual into $\lambda\mathbb E[f(W+1)-f(W^s)]$. Telescoping over
	integer arguments gives the displayed total-variation bound. In the stationary extension, the nearest flags on both
	sides are almost surely at finite distances. Restricting the coupled region to the actual observation interval
	only removes indicators, so no endpoint correction is missing. This proves
	Eq.~\eqref{eq:oct18-poisson-general}.
	\end{proof}
	The flags and resets are auxiliary variables used only in the proof.
	The receiver cannot observe them or reset the source.

\subsection{A bound on the map for rejected events}
We use the unnormalized Choi matrix $J(\mathcal T)=\sum_{a,b=1}^{d}|a\rangle\langle
b|\otimes\mathcal T(|a\rangle\langle b|)$. For $\pi=I/d$, suppose
$\|\Phi^KX\|_2\le s_K\|X\|_2$ on trace-zero matrices. Choi reshuffling preserves the
Frobenius norm, and the superoperator $\Phi^K-D$ has at most $d^2-1$ singular values, each
at most $s_K$. Since $J(D)=I_{d^2}/d$,
\[
\Phi^K\gecp[1-d\sqrt{d^2-1}s_K]D.
\]
The deleted-event map is CP by the telescoping identity
\[
C_K=\Phi^K-B^K=\sum_{j=0}^{K-1}\Phi^{K-1-j}AB^j.
\]
Because $A^*(I)\le d\mu I$ and $B$ is trace nonincreasing,
$\Tr J(C_K)=\Tr C_K(I)\le d^2K\mu$. Consequently
\begin{equation}
	B^K\gecp[1-d\sqrt{d^2-1}s_K-d^3K\mu]D.
	\label{eq:oct18-complement-criterion}
\end{equation}
For a general full-rank $\pi$, define
\[
\delta_K=\|(I\otimes\pi^{-1/2})[J(\Phi^K)-I\otimes\pi]
(I\otimes\pi^{-1/2})\|.
\]
Then $\Phi^K\gecp(1-\delta_K)D_\pi$ and $\Tr J(C_K)\le dK\mu/m$. It follows that
\begin{equation}
	B^K\gecp[1-\delta_K-dK\mu/m^2]D_\pi.
\end{equation}
For a fixed primitive channel, $\delta_K\to0$. We can choose $K$ first, then require the event to be sufficiently rare. Primitivity alone does not make an arbitrary event rare enough. Uniform families need a common positive lower bound on the smallest eigenvalue of $\pi$ and common $K,c$. For an initial state $\rho$, include $b$ relaxation blocks in the observation time. State-to-output contractivity adds at most $\frac12\|\Phi^b\rho-\pi\|_1$ to the total-variation bound for the entire subsequent record. This cost is paid once, not once per later block.

\subsection{Initial relaxation and the constants used in the Letter}
For this ring the physical dissipator contracts traceless matrices in the Hilbert--Schmidt norm
at rate $2-\sqrt2>1/2$, uniformly in $h$. Fifty five-bin blocks, each of duration $1/5$,
span a time of ten, so $s_K\le \e^{-5}$. With $\mu_h<10^{-10}$ from
Sec.~\ref{sec:finite-example}, Eq.~\eqref{eq:oct18-complement-criterion} gives
\begin{equation}
	(\Phi_h-B_h^{\rm event})^{50}\gecp
	\left[1-\frac{64}{148}-\frac{512\cdot50}{10^{10}}\right]D>\tfrac12D.
\end{equation}
Here $8\sqrt{63}<64$ and $\e^5>148$, so $d\sqrt{d^2-1}s_K\le8\sqrt{63}\,\e^{-5}\le64/148$.
Substituting $c=1/2$, $m=1/8$, and $K=50$ in
Eq.~\eqref{eq:oct18-poisson-general} gives $9\cdot50\cdot(4-1)/(1/2)=2700$. The
counted initial interval contributes
\begin{equation}
	\tfrac12\|\e^{20\mathcal L_h}P_4-I/8\|_1
	\le(\sqrt7/2)\e^{-10}<3/40000.
\end{equation}
The change in the entire subsequent count law is bounded by the same trace distance,
independently of its duration. The resulting error is $3/40000+2700\mu_h$, plus the
explicitly bounded displacement of any rounded probability center. The map decomposition
introduces no observed reset or feedback.